\documentclass[12pt]{article}
\usepackage{amsmath,amssymb,epsfig}
\usepackage{cite}

\usepackage{color}
\input{colordvi.tex}
\def\unit{{\relax{\rm 1\kern-.26em I}}}

\newcommand{\tr}{{\rm Tr}}

\makeatletter

\renewcommand\section{\@startsection {section}{1}{\z@}%
                                   {-3.5ex \@plus -1ex \@minus -.2ex}%
                                   {2.3ex \@plus.2ex}%
                                   {\normalfont\large\bfseries}}

\renewcommand\subsection{\@startsection{subsection}{2}{\z@}%
                                     {-3.25ex\@plus -1ex \@minus -.2ex}%
                                     {1.5ex \@plus .2ex}%
                                     {\normalfont\normalsize\bfseries}}

\makeatother

\begin{document}

\baselineskip=18pt  
\numberwithin{equation}{section}  
\allowdisplaybreaks  



%
%


\thispagestyle{empty}

\vspace*{-2cm}
\begin{flushright}
\end{flushright}

\begin{flushright}

\end{flushright}

\begin{center}

\vspace{1.4cm}

\vspace{0cm}
{\bf \Large Diphoton Resonance from a New Strong Force
}
\vspace*{1.3cm}

{\bf 
Howard Georgi and Yuichiro Nakai} \\
\vspace*{0.5cm}

{\it Department of Physics, Harvard University, Cambridge, MA 02138, USA}\\

\vspace*{0.5cm}

\end{center}

\vspace{0.5cm} \centerline{\bf Abstract} \vspace*{0.5cm}

We explore a ``partial unification'' model that could explain the diphoton 
event excess around $750 \, \rm GeV$ recently reported by the LHC experiments.
A new strong gauge group is combined with the ordinary color and
hypercharge gauge groups. The VEV responsible for the combination is of the
order of the $SU(2)\times U(1)$ breaking scale, but the coupling of the new
physics to standard model particles is suppressed by the strong
interaction of the new gauge group.
This simple extension of the standard model has a rich phenomenology,
including composite particles of the new confining gauge interaction, a
coloron and a $Z'$ which are 
rather weakly coupled to standard model particles, and
massive vector bosons charged 
under both the ordinary color and hypercharge gauge groups and the new
strong gauge group. 
The new scalar glueball could have mass of around $750 \, \rm GeV$,
be produced by gluon fusion and decay into two photons, both through
loops of the new massive vector bosons.
The simplest version of the model has some issues: the massive vector
bosons are stable and the coloron and the $Z'$  are
strongly constrained by search data. An
extension of the model to include additional fermions with the new gauge
coupling, though not as simple 
and elegant, can address both
issues and more. 
It allows the massive vector boson to decay
into a colorless, neutral state that could be a candidate of the dark
matter.  And the coloron and $Z'$ can decay dominantly into the new
fermions, completely changing the search bounds.  In addition, $SU(N)$
fermions below the symmetry breaking scale make it more plausible that the
lightest glueball is at $750$~GeV. 
If the massive vector bosons are still long-lived,
they could form new bound states, ``vector bosoniums'' with additional
interesting phenomenology.  Whatever becomes of the $750$~GeV diphoton
excess, the model is an unusual example of how new physics at small scales could be
hidden by strong interactions.

\newpage
\setcounter{page}{1} 



\section{Introduction\label{sec:intro}}

Recently, the ATLAS and CMS collaborations have reported
an event excess in diphoton invariant mass distribution of around $750 \, \rm GeV$
\cite{ATLAS,CMS:2016owr}.
If this excess is real and comes from a new scalar (or pseudoscalar) particle $S$
decaying into two photons, the relatively large production cross section
times branching ratio 
$\sigma (pp \rightarrow S)  {\rm Br} (S \rightarrow \gamma \gamma)$
required to fit the data suggests
a nonperturbatively large coupling of $S$ with electrically charged particles and
hence the existence of new strong dynamics (see e.g. Ref~\cite{Harigaya:2015ezk,Nakai:2015ptz,Franceschini:2015kwy,Craig:2015lra}).

In this paper, we report on an exercise in model building that is
loosely
motivated by the diphoton excess.
We consider the possibility that this is a real effect
from a previously unobserved strong gauge interaction.
To connect the new strong dynamics with the diphoton excess,
we pursue a ``partial unification'' scenario
in which a part of the ordinary color and hypercharge gauge groups
and the new strong gauge group are 
combined near the $SU(2)\times U(1)$-breaking scale of $250$~GeV.
We do not pretend that
the structures we explore here are in any way unique and certainly not that
they are well-motivated. But we do believe that it
is interesting to build a very explicit and minimal model of such a scenario.
This very simple extension of the standard model has a rich phenomenology
at the TeV scale, 
including massive vector bosons which we call $X',\bar{X}'$,
charged
under both the ordinary color and hypercharge gauge groups and the new strong
gauge group.
The model also contains color octet vector bosons (colorons
\cite{Chivukula:1996yr,Simmons:1996fz,Bai:2010dj}) and a $Z'$ (see for
example \cite{Langacker:2008yv}) both of which are rather weakly coupled to
standard model particles, and color singlet and octet scalars (see for example
\cite{Bai:2010dj}).  

The lightest glueball associated with the new strong gauge interaction
could be
a scalar particle with mass of around $750 \, \rm GeV$.\footnote{
Ref \cite{Harigaya:2015ezk} has discussed a glueball explanation for the diphoton excess in a different model.}
This is natural in our model
because all the other new states have
masses that scale with the partial unification scale.
The scale of confinement
after the gauge symmetry breaking is generically smaller.
The new scalar glueball is efficiently produced by gluon fusion and decays
into two photons through 
loops of the new massive vector bosons $X',\bar{X}'$.

The simplest version of the model has some issues: the massive vector
bosons are stable and the coloron and  $Z'$ are
strongly constrained by search data. An
extension of the model to include additional fermions, though not as simple
and elegant, can address both
issues and more.  We show how this can 
allow the $X'$ boson to decay into quarks and antiquarks 
plus a colorless, neutral state
that could be an 
unusual dark matter candidate. 
The decays of the coloron and $Z'$ into pairs of the new fermions are
important in evading search constraints.  There are many possible extensions of this
kind, depending on the details of the partial unification. In the
particular example we discuss in detail, the model contains one or more charge $5/3$
quarks and neutral fermions with the new strong interaction  
in a fairly narrow mass range from above $375$~GeV (half the mass of the
lightest glueball) to less than half the mass of the coloron.   
In addition, the $SU(N)$
fermions below the symmetry breaking scale make it more plausible that the
lightest glueball is at $750$~GeV. 

Apart from the mass and decay constant of the lightest glueball, which we
take from lattice gauge 
theory studies~\cite{Morningstar:1999rf, Chen:2005mg, Jaffe:1985qp}, the
relevant spectrum and interactions in our model can be calculated
perturbatively in some regions of the parameter space.  However, as we will
see, to explain the data, we will be pushed into a region of parameter space
where the new gauge coupling is rather large, so some of our estimates may
be only rough approximations, and indeed, we can't even be sure that the
relevant symmetry breaking takes place as the perturbation theory suggests.
Conversely, if the excess persists, and some scenario like this turns out
to be the right explanation, we will learn a tremendous amount about strong
gauge interactions that are very different from QCD.

The rest of the paper is organized as follows.
In section~\ref{sec:model}, we present our model and analyze the mass spectra.
We discuss some of the experimental constraints on the model in
section~\ref{sub:constraints}. 
In section~\ref{sec:glueballs}, we discuss the possibility that
the lightest glueball associated with the new strong gauge interaction that
is partially unified with color $SU(3)$ at a relatively low scale
could explain the diphoton excess. 
In section~\ref{sec:xdecay}, we add additional fermions to the model
transforming under the new gauge interaction. 

If the $X', \bar{X}'$ gauge bosons are still long-lived,
they form new bound states, ``vector bosoniums.''
Detailed phenomenology of the vector bosoniums is left for a future study.
Various details including group theory notation and identities and some of
the interactions in the model 
are relegated to appendices.


\section{Partial unification\label{sec:model}}

Here we discuss a minimal extension of the standard model in which a part of the
color $SU(3)$ and the hypercharge $U(1)$ resides in an extended gauge group.
The $U(1)$ normalization is important for determining the electric charge of the new massive vector bosons.
In this section, we describe the symmetry breaking in detail and
analyze the mass spectra of the scalar fields and the vector bosons.

\subsection{The $SU(N+3)$ model\label{subsec:sun}}

\renewcommand{\arraystretch}{1.3}
\begin{table}[!t]
\begin{center}
\begin{tabular}{c|ccccc}
 & $SU(N+3)_{H}$ & $SU(3)_{C'}$ & $SU(2)_{L}$ & $U(1)_{Y'}$ 
 \\
 \hline
  $\xi$  &  $\mathbf{N+3}$ & $\mathbf {\bar 3}$ & $\mathbf  1$ & $-\frac{Nq}{N+3}$ \\
\end{tabular}
\end{center}
\caption{The charge assignment of the $\xi$ field.
The $U(1)_{Y'}$ charge of $\xi$ is explained in the main text.}
\label{tab:sunmodel}
\end{table}
\renewcommand{\arraystretch}{1}
\renewcommand{\arraystretch}{1.3}
\begin{table}[!t]
\begin{center}
\vspace{0.3cm}
\begin{tabular}{c|ccc}
 & Gauge field & Gauge coupling & Generator
 \\
 \hline
  $SU(N+3)_H$  &  $H_\mu^A$ & $g_H$ & $T^A$ \\
  $SU(3)_{C'}$  &  $A{^\prime}^a_\mu$ & $g'_3$ & $T^a$ \\
  $SU(2)_L$  &  $W_\mu^\alpha$ & $g_2$ & $T^\alpha$ \\
  $U(1)_{Y'}$  &  $B'_\mu$ & $g'_Y$ & $S_{Y'}$ \\
\end{tabular}
\end{center}
\caption{The names of gauge fields, gauge couplings and generators of the model.
Here, $A = 1, \cdots, (N+3)^2 -1$, $a = 1, \cdots , 8$ and $\alpha = 1,2,3$.
More group theory notation and identities are summarized in appendix A.}
\label{tab:gauge}
\end{table}
\renewcommand{\arraystretch}{1}

We introduce a new $SU(N+3)_H$ gauge theory with a complex scalar $\xi$
which is charged under both the $SU(N+3)_H$ gauge group and
the (would-be) standard model gauge groups $SU(3)_{C'} \times U(1)_{Y'}$.
The charge assignment is shown in Table~\ref{tab:sunmodel}.  Thus the $\xi$
transforms like $(N+3,\bar 3)_{-Nq/(N+3)}$ under $SU(N+3)_H\times
SU(3)_{C'} \times U(1)_{Y'}$ and it is convenient to represent it as an
$(N+3)\times3$ matrix.
The names of gauge fields, gauge couplings and generators are summarized in Table~\ref{tab:gauge}.
The ordinary standard model particles have the conventional 
charges under the $SU(3)_{C'} \times SU(2)_{L} \times U(1)_{Y'}$.
We can also introduce new matter fermions charged under the $SU(N+3)_H$
gauge group, 
which have an interesting role in the massive gauge boson decay.  This will
be discussed in section~\ref{sec:xdecay}.
The most general potential involving only the scalar $\xi$ can be written as
\begin{equation}
\begin{split}
\\[-2.5ex]
V_\xi \, = \, &\frac{1}{4} \lambda_1 \left(\tr ( \xi^{\dagger} \xi) - 3 a^2 \right)^2
+ \frac{1}{2} \lambda_2 \, \tr \left( \xi^{\dagger} \xi - a^2 I_3 \right)^2
\label{coloredpotential} \, , \\[1ex]
\end{split}
\end{equation}
where $I_3$ is the $3 \times 3$ identity matrix,
$\lambda_1$, $\lambda_2$ are dimensionless parameters
and $a$ is a mass parameter.

For the range of parameters
\begin{equation}
\lambda_2>0 \, , 
\quad\quad
\lambda_1>-\frac{2}{3}\lambda_2 \, ,
\label{range}
\end{equation}
the potential \eqref{coloredpotential} is minimized when
some of the components in $\xi$
get nonzero vacuum expectation values.
The vev can be put in the following form
\begin{equation}
\begin{split}
\\[-2.5ex]
\langle \xi \rangle = 
\begin{pmatrix}
a &0 &  0 \\
0 & a &0 \\
0 & 0 & a  \\
0&0&0\\
\vdots&\vdots&\vdots
\end{pmatrix}  \, , \label{vev} \\[1ex]
\end{split}
\end{equation}
and the gauge groups $SU(N+3)_{H} \times SU(3)_{C'} \times SU(2)_{L} \times U(1)_{Y'}$ are  spontaneously broken to
$SU(N)_{H} \times SU(3)_{C} \times SU(2)_{L} \times U(1)_{Y}$.
Below the scale $a$, the gauge structure is just the conventional
standard model with an additional $SU(N)$ gauge group that does not couple
to the standard model particles.  Thus for a large $a$ the gauge couplings of the 
$SU(3)_{C} \times SU(2)_{L} \times U(1)_{Y}$ would be just the standard model
couplings to a good approximation.  However, we will see that this is not
an interesting limit.  Instead, we will be interested in $a$ of the order
of (or even smaller than) the $SU(2)\times U(1)$ breaking scale
$v\approx250$~GeV.  We will try to convince you of the somewhat surprising
statement that such a low value of $a$ is not ruled out by current data.
Roughly speaking, this works because the heavy gauge boson masses are of the order
of $a$ times a large coupling $g_H$, and in many cases can be integrated out as
if $a$ were large.  In general, this is a dangerous procedure, because the
large coupling can appear in the numerator and spoil decoupling.  But here
is it often OK, because there are no direct order-$g_H$ couplings to
the standard model particles.

The model contains massive particles whose masses scale with $a$.
Corresponding to the broken symmetries, there are massive $X', \bar{X}'$ gauge bosons, charged
under both $SU(3)_C$, $U(1)_Y$ and the new $SU(N)_H$, as well as the $Z'$ and 
the color octet $G'$ vector bosons.
The $G'$ gauge boson is also called as the coloron.
The charge assignments of the massive vector bosons are summarized in
Table~\ref{tab:sunvector}. 
In this mass range, there are also massive scalars, $G_O$ and $G_S$
transforming like an octet and singlet respectively under the color $SU(3)_C$.
Their mass spectra are analyzed below.

\renewcommand{\arraystretch}{1.3}
\begin{table}[!t]
\begin{center}
\begin{tabular}{c|ccccc}
 & $SU(N)_{H}$ & $SU(3)_{C}$ & $SU(2)_{L}$ & $U(1)_{Y}$ 
 \\
 \hline
  $X'$  &  $\mathbf{\overline{N}}$ & $\mathbf 3$ & $\mathbf  1$ & $q$ \\
  $\bar{X}'$ & $\mathbf{{N}}$ & $\mathbf{\bar{3}}$ & $\mathbf  1$ & $-q$  \\
  $Z'$ & $\mathbf{{1}}$ & $\mathbf 1$ & $\mathbf  1$ & $0$  \\
  $G'$ & $\mathbf{{1}}$ & $\mathbf 8$ & $\mathbf  1$ & $0$  \\
\end{tabular}
\end{center}
\caption{The charge assignments of the massive vector bosons.}
\label{tab:sunvector}
\end{table}
\renewcommand{\arraystretch}{1}

\subsection{Gauge couplings\label{subsec:u1norm}}

After the symmetry breaking, the ordinary $SU(3)_C$, $U(1)_Y$ gauge groups are given by
combinations of the $SU(N+3)_H$ gauge group
and the $SU(3)_{C'}$, $U(1)_{Y'}$ gauge groups.
The ordinary massless gluons and their gauge coupling $g_s$ are given by the following relations,
\begin{equation}
\begin{split}
\\[-2.5ex]
A^{a}_\mu = \frac{g'_3 H_\mu^a + g_H  A{^\prime}^a_\mu}{\sqrt{g_H^2 + (g'_3)^2} } \, , \qquad
\frac{1}{g_s^2} = \frac{1}{(g_{H})^2} + \frac{1}{(g'_3)^2}  \, , \label{colorrelation} \\[1ex]
\end{split}
\end{equation}
where $g_H$ and $g'_3$ are the gauge couplings of the $SU(N+3)_H$ and $SU(3)_{C'}$ gauge groups respectively.
The field $H^a_\mu$ ($a = 1, \cdots , 8$) is the $SU(3)$ part of the $SU(N+3)_H$ gauge field $H_\mu^A$
($A = 1, \cdots, (N+3)^2 -1$).

We next consider the $U(1)$ normalization.
The $U(1)_{Y'}$ charge of $\xi$ is given by $-N/(N+3)$ times the $U(1)_Y$
charge of the $X'$ gauge boson, 
which we call $q$.
The $U(1)$ subgroup of $SU(N+3)_H$ commuting with the $SU(3)$ and $SU(N)$
subgroups, is generated by
\begin{equation}
\begin{split}
\\[-2.5ex]
S_H \equiv \frac{q}{N+3} \begin{pmatrix}
N \, I_3 & 0 \\
0 & -3 \, I_N \\
\end{pmatrix}   \, , \qquad
\left[ S_H , T_{3 \times N} \right] = 0 \, , \qquad T_{3 \times N} = \begin{pmatrix}
T_3 & 0 \\
0 & T_N \\
\end{pmatrix}   \, , \\[1ex]
\end{split}
\label{qunify}
\end{equation}
which is normalized so that the $U(1)_Y$ charge of the standard model is given by
\begin{equation}
\begin{split}
\\[-2.5ex]
S_Y = S_H + S_{Y'}  \, . \\[1ex]
\end{split}
\end{equation}
In this case, we correctly obtain $S_Y \langle \xi \rangle = 0$, which means $S_Y$ is not broken in the effective theory between
the scale $a$ and the Higgs vev.
On the other hand, the properly normalized generator of the $U(1)$ subgroup is
\begin{equation}
\begin{split}
\\[-2.5ex]
\tilde{S}_H = k S_H \, , \qquad k = \frac{\sqrt{N+3}}{q \sqrt{6N}} \, . \\[1ex]
\end{split}
\end{equation}
Note that because the $SU(N+3)_H$ does not involve the electroweak $SU(2)_L$,
there is no constraint on the $X'$ charge $q$ from the structure of the
electroweak interactions.  However, if we require that states that are
singlets under the color $SU(3)_C$ and the confining $SU(N)_H$ are integrally charged,
then we have the constraint
\begin{equation}
\begin{split}
\\[-2.5ex]
&q = \frac{3j-1}{3N} \quad \text{for integer $j$ if $N$ mod 3 = 1} \, , \\[1ex]
&q = \frac{3j+1}{3N} \quad \text{for integer $j$ if $N$ mod 3 = 2} \, , \\[1ex]
&q = \frac{j}{N} \quad \text{for integer $j$ if $N$ mod 3 = 0} \, . \\[1ex]
\end{split}
\end{equation}
Another constraint on $q$ is discussed below.

The ordinary massless hypercharge gauge field $B_\mu$ and its gauge coupling $g_Y$ are given by
\begin{equation}
\begin{split}
\\[-2.5ex]
B_\mu  = \frac{ g'_Y B''_\mu +
 g''_Y B'_\mu}{\sqrt{ (g'_Y)^2 + (g''_Y)^2} } \, , \qquad
\frac{1}{g_Y^2} =  \frac{1}{ (g'_Y)^2} +  \frac{1}{(g''_Y)^2} \, , \label{hyperchargerelation} \\[1ex]
\end{split}
\end{equation}
where 
\begin{equation}
g''_Y = k g_H \, , \qquad k= \frac{\sqrt{N+3}}{q \sqrt{6N}} \, ,
\label{k}
\end{equation}
and $B''_\mu$ is the (properly normalized) $U(1)$ part of the $SU(N+3)_H$
gauge field.  
Because the low energy theory is identical to the standard model as $a\to\infty$, this
implies that to leading order in $(v/a)^2$ ($v$ is the Higgs vev)
\begin{equation}
\begin{split}
\\[-2.5ex]
\frac{1}{\sqrt{1/ (k g_H)^2 + 1 / (g'_Y)^2}} \simeq \frac{e}{ \cos \theta_W}\, . \\[1ex]
\end{split}
\end{equation}
Here, $\sin^2 \theta_W = 0.23$ is the weak mixing angle and $e$ is the electromagnetic gauge coupling.
Solving this equation for $g'_Y$, we obtain
\begin{equation}
\begin{split}
\\[-2.5ex]
g'_Y \simeq \frac{1}{\sqrt{\frac{\cos^2 \theta_W}{e^2} - \frac{6Nq^2}{(N+3) g_H^2}}}\, , \label{gYprime} \\[1ex]
\end{split}
\end{equation}
which implies that $q$ cannot be too large.

\subsection{Massive vector bosons\label{subsec:xs}}

We here analyze the mass spectrum of the $G'$, $Z'$ and $X'$, $\bar{X}'$ gauge bosons and some of their interactions.
From the covariant derivative of the scalar $\xi$ which has the vev \eqref{vev},
the coloron $G'$ and the massive vector boson corresponding to the broken $U(1)$ are given by the linear combinations,
\begin{equation}
\begin{split}
\\[-2.5ex]
G{^\prime}^a_\mu = \frac{g_H H_\mu^a - g'_3 A{^\prime}^a_\mu}{\sqrt{g_H^2 + (g'_3)^2} } \, ,
\qquad B^-_\mu = \frac{ g''_Y B''_\mu -  g'_Y B'_\mu}{\sqrt{ (g'_Y)^2 +  (g''_Y)^2} }  \, . \\[1ex]
\end{split}
\end{equation}
The vector boson masses are
\begin{equation}
\begin{split}
\\[-2.5ex]
&m_{G'}^2 = a^2 \left(g_H^2 + (g'_3)^2 \right) \, , \\[1ex]
&m_{B^-}^2 = 6 \left( \frac{Nq}{N+3} \right)^2 a^2 \left( (g'_Y)^2 +  (g''_Y)^2 \right) \, , \\[1ex]
&m_{X'}^2 =  g_H^2a^2 \, . \\[1ex]
\end{split}
\end{equation}
Note that the coloron is always heavier than the $X'$, $\bar{X}'$ gauge bosons.

After $SU(2)\times U(1)$ breaking,
the massless photon field is the linear combination,
\begin{equation}
\begin{split}
\\[-2.5ex]
{A}_\mu = \frac{g_Y W_\mu^3 + g_2 B_\mu}{\sqrt{ (g_2)^2 +  (g_Y)^2}} \, . \\[1ex]
\end{split}
\end{equation}
The two massive eigenstates are given by
\begin{equation}
\begin{split}
\\[-2.5ex]
Z_\mu = \cos \omega \hat{Z}_\mu + \sin \omega B_\mu^-  \, , \qquad
Z'_\mu = - \sin \omega \hat{Z}_\mu + \cos \omega B_\mu^- \, , \label{omega} \\[1ex]
\end{split}
\end{equation}
where
\begin{equation}
\begin{split}
\\[-2.5ex]
\hat{Z}_\mu = \frac{g_2 W_\mu^3 - g_Y B_\mu}{\sqrt{ (g_2)^2 +  (g_Y)^2}} \, , \qquad
\tan 2 \omega = -2 \frac{\delta \hat{m}^2 }{\hat{m}^2_{B^-} - \hat{m}_Z^2} \, , \label{Zhat} \\[1ex]
\end{split}
\end{equation}
and
\begin{equation}
\begin{split}
\\[-2.5ex]
&\hat{m}_Z^2 = \frac{1}{4} v^2 \left((g_2)^2 + (g_Y)^2 \right)  \, , \\[1ex]
&\hat{m}_{B^-}^2 = \frac{1}{4} v^2 \frac{(g'_Y)^4 }{(g'_Y)^2 +  (g''_Y)^2}
+ m^2_{B^-}  \, , \\[1ex]
&\delta \hat{m}^2 = \frac{1}{4} v^2 (g'_Y)^2 \frac{\sqrt{(g_2)^2 +  (g_Y)^2}}{\sqrt{(g'_Y)^2 +  (g''_Y)^2}}  \, . \label{ZmassLagrangian} \\[1ex]
\end{split}
\end{equation}
The eigenvalues are
\begin{equation}
\begin{split}
\\[-2.5ex]
&{m}_Z^2 = \frac{1}{2} \left( \hat{m}_Z^2 + \hat{m}_{B^-}^2 -
\sqrt{\left( \hat{m}_Z^2 - \hat{m}_{B^-}^2 \right)^2 + 4 \delta \hat{m}^4 }  \right) \, , \\[1ex]
&{m}_{Z'}^2 = \frac{1}{2} \left( \hat{m}_Z^2 + \hat{m}_{B^-}^2 +
\sqrt{\left( \hat{m}_Z^2 - \hat{m}_{B^-}^2 \right)^2 + 4 \delta \hat{m}^4 }  \right) \, . \\[1ex]
\end{split}
\end{equation}

We now summarize the interactions of the massive gauge bosons with the standard model fermion $f$ for later purposes.
The $X', \bar{X}'$ gauge bosons do not couple to the standard model fermion at tree level.
The coloron interaction with the standard model fermion is
\begin{equation}
\begin{split}
\\[-2.5ex]
\mathcal{L} \, \supset \, 
- \frac{(g'_3)^2}{\sqrt{g_H^2 + (g'_3)^2}} \, \bar{f} \gamma^\mu T^a f \, G{^\prime}^{ a }_\mu \, . \label{coloronint}  \\[1ex]
\end{split}
\end{equation}
The important point is that the coupling is small when the $g_H$ coupling
is large.  This will be the interesting region for our analysis. 
In this region, $g'_3\approx g_s$ by the relation \eqref{colorrelation}.

The $Z'$ couplings are more complicated because the $SU(2)$ symmetry
breaking scale $v$ is important.  Even
though we will keep the new symmetry breaking scale, $a$, of the same order
as $v$, because the strong $SU(N+3)$ group is not directly coupled to
standard model particles, we will be able to expand quantities in $1/g_H$
to simplify our expressions and understand what is going on. 
At leading order in $1/g_H$, the masses satisfy
\begin{equation}
g_Ha\approx m_{G'}\approx m_{X'}\approx \sqrt{\frac{N+3}{N}}m_{Z'}
\label{largeghmasses}
\end{equation}
and the $Z'$ interaction is
\begin{equation}
\begin{split}
\\[-2.5ex]
\mathcal{L} \, \supset \, 
- \sqrt{\frac{6Nq}{(N+3)}}\frac{(g'_Y)^2}{g_H} 
\left( Y^{f}_L \bar{f}_L \gamma^\mu f_L 
+ Y^{f}_R \bar{f}_R \gamma^\mu f_R \right) Z'_\mu \, . \label{Zprimeint}  \\[1ex]
\end{split}
\end{equation}
Here, $Y^f_{L,R}$ are the hypercharges of the left and right-handed fermions $f_{L,R}$.
Again the interaction is suppressed when the $g_H$ coupling is large
and $g'_Y\approx e/\cos\theta_W$ by the relation \eqref{hyperchargerelation}.

\subsection{Scalar mass spectrum\label{subsec:scalars}}

The scalar $\xi$ has $6(N+3)$ (real) degrees of freedom.
Here, $8+1+6N$ of them are unphysical
Nambu-Goldstone modes eaten in the symmetry breaking.
Thus there are $9$ physical degrees of freedom.
The potential of the scalar sector is given by \eqref{coloredpotential}
plus terms involving the standard model Higgs $\phi$,
\begin{equation}
\begin{split}
\\[-2.5ex]
V_{\rm Higgs} \, = \,
\frac{1}{4} \lambda_3 \left(  \phi^{\dagger} \phi -  \frac{v^2}{2} \right)^2
+ \lambda_4 \left(  \phi^{\dagger} \phi -  \frac{v^2}{2} \right)
\left( \tr ( \xi^{\dagger} \xi) - 3 a^2  \right) \, , \label{scalarpotential} \\[1ex]
\end{split}
\end{equation}
where $\lambda_3$ and $\lambda_4$ are dimensionless coupling constants.
To analyze the mass spectrum of the physical modes, we now take unitary gauge,
\begin{equation}
\begin{split}
\\[-2.5ex]
\xi = 
\begin{pmatrix}
a I_3 + \chi/\sqrt{2} \\
0
\end{pmatrix}  \, , \qquad \phi = \frac{1}{\sqrt{2}}
\begin{pmatrix}
0 \\
v + h
\end{pmatrix}  
\, , \qquad \mbox{where $\chi^\dagger=\chi$}
\, .
\label{unitary} \\[1ex]
\end{split}
\end{equation}
The trace and traceless parts of $\chi$ are singlet and octet under the color $SU(3)_C$ respectively.
Properly normalizing the kinetic terms, the color octet/singlet scalars are
written (using the Gell-Mann matrices $\lambda^a$) as
\begin{equation}
\begin{split}
\\[-2.5ex]
G_O^a=\tr\left(\lambda^a \chi\right)
 \, ,
\qquad G_S \equiv \sqrt{\frac{2}{3}}  \tr \left( \chi\right)
\end{split}
\end{equation}
Then, the mass of the octet scalar $G_O$ is given by
\begin{equation}
\begin{split}
\\[-2.5ex]
m_{G_O}^2 = 2\lambda_2 a^2 \, . \\[1ex]
\end{split}
\end{equation}
Due to the second term of \eqref{scalarpotential}, the singlet component $G_S$ mixes with the Higgs field $h$.
The mass eigenstates are
\begin{equation}
\begin{split}
\\[-2.5ex]
\phi_1 = h \cos \theta_h + G_S \sin \theta_h \, ,
\qquad \phi_2 = - h \sin \theta_h + G_S \cos \theta_h \, . \\[1ex]
\end{split}
\end{equation}
The mixing angle $\theta_h$ is given by
\begin{equation}
\begin{split}
\\[-2.5ex]
\tan 2 \theta_h = - \frac{2 \sqrt{6} \lambda_4 v a}{m_{G_S}^2 - m_h^2} \, , \\[1ex]
\end{split}
\end{equation}
where
\begin{equation}
\begin{split}
\\[-2.5ex]
m_{G_S}^2 = \left( 3 \lambda_1 + \frac{2}{3} \lambda_2 \right) a^2 \, , \qquad
m_h^2 = \frac{1}{2} \lambda_3 v^2 \, . \\[1ex]
\end{split}
\end{equation}
The eigenvalues are
\begin{equation}
\begin{split}
\\[-2.5ex]
m_{\phi_1}^2 = \frac{1}{2} \left( m_h^2 + m_{G_S}^2 \right) 
- \frac{1}{2} \sqrt{\left( - m_h^2 + m_{G_S}^2 \right)^2 + 24 \lambda_4^2 v^2 a^2 }\, , \\[2ex]
m_{\phi_2}^2 = \frac{1}{2} \left( m_h^2 + m_{G_S}^2 \right) 
+ \frac{1}{2} \sqrt{\left( - m_h^2 + m_{G_S}^2 \right)^2 + 24 \lambda_4^2 v^2 a^2 }\, . \\[1ex]
\end{split}
\end{equation}
The mass of the lighter eigenstate $m_{\phi_1}$ gives the physical Higgs
boson mass, $m_{\phi_1} \simeq 125 \, \rm GeV$.

\section{Experimental constraints\label{sub:constraints}}

In this section, we discuss the experimental constraints on the new
parameters that we have introduced in our extension of the standard
model. The possible constraints are of three kinds.  There are constraints from
precise tests of the standard model at low energies.
There are ``conpositeness'' constraints on the virtual effects of the new particles.
In addition, there are bounds
from direct searches for the new particles  in our model,
in particular the lower bounds on the $Z'$ mass and the coloron mass.

\subsection{Electroweak precision tests\label{subsec:sandt}}

For a sufficiently large $a$, the low-energy interactions of the standard
model particles are indistinguishable from their standard model limits. 
But our $a$ will not be large, so 
precise tests of the standard model create interesting constraints.
Let us consider the $U(1)$ part of the model,
\begin{equation}
\begin{split}
\\[-2.5ex]
\mathcal{L} &\, \supset \, 
- \frac{1}{4} B'_{\mu\nu} {B'}^{\mu\nu} - \frac{1}{4} B''_{\mu\nu} {B''}^{\mu\nu} 
+ \frac{1}{2} m_{B^-}^2 B_\mu^- B^{-\mu} \, . \\[1ex]
\end{split}
\end{equation}
Here, $B'_{\mu\nu}$ and $B''_{\mu\nu}$ are the field strengths of the
$B'_\mu$ and $B''_\mu$ gauge fields respectively. 
The $B'_\mu$ field couples to the usual standard model fields with the
gauge coupling $g'_Y$. 
We can integrate out the heavy mode at tree level by solving 
the equation of motion for the $B''_\mu$ field, given by
\begin{equation}
\begin{split}
\\[-2.5ex]
\partial^\nu B''_{\mu\nu}  - m_{B''}^2 B''_\mu =  - \frac{g'_Y}{g''_Y}  m_{B''}^2 B'_\mu \, , \qquad
m_{B''}^2 \equiv \left( \frac{(g''_Y)^2}{(g'_Y)^2 + (g''_Y)^2} \right) m_{B^-}^2 \, . \\[1ex]
\end{split}
\end{equation}
We have defined a handy parameter $m_{B''}^2$ which is not a physical mass.
This equation of motion has a solution,
\begin{equation}
\begin{split}
\\[-2.5ex]
B''_{\mu\nu}  = \frac{g'_Y}{g''_Y} \frac{\delta^\nu_\mu + \partial_\mu \partial^\nu / m_{B''}^2}
{1 + \partial^2 / m_{B''}^2} B'_\nu \, . \\[1ex]
\end{split}
\end{equation}
Thus the Lagrangian after integrating out $B''_\mu$ at tree level is given by
\begin{equation}
\begin{split}
\\[-2.5ex]
\mathcal{L}_{\rm eff} &\, \supset \, 
- \frac{1}{4} B'_{\mu\nu} {B'}^{\mu\nu} - \frac{1}{4} \left( \frac{g'_Y}{g''_Y} \right)^2
B'_{\mu\nu} \left( \frac{1}{1+ \partial^2 / m_{B''}^2} \right) {B'}^{\mu\nu}  \\[1.5ex]
&\, = \, - \frac{1}{4} \frac{(g'_Y)^2 + (g''_Y)^2}{(g''_Y)^2}B'_{\mu\nu} {B'}^{\mu\nu} + \frac{1}{4} \left( \frac{g'_Y}{g''_Y} \right)^2
B'_{\mu\nu} \left( \frac{\partial^2}{m_{B''}^2} \right) {B'}^{\mu\nu} + \cdots \, . \\[1ex]
\end{split}
\end{equation}
We have omitted to write irrelevant dimension eight and higher operators.
Correctly normalizing the kinetic term as in \eqref{hyperchargerelation},
we obtain
\begin{equation}
\begin{split}
\\[-2.5ex]
\mathcal{L}_{\rm eff} &\, \supset \, 
- \frac{1}{4} B_{\mu\nu} B^{\mu\nu} 
- \frac{1}{4} \left( \frac{g'_Y}{g''_Y} \right)^2
\frac{1}{m_{B^-}^2} (\partial_\rho B^{\mu\nu} )^2 + \cdots \, , \\[1ex]
\end{split}
\end{equation}
where $B_\mu$ is the ordinary hypercharge gauge field which couples to the standard model fields with the gauge coupling $g_Y$.
Note that there are no effective operators to give the $S$, $T$ and $U$ parameters
\cite{Peskin:1991sw}.
However, the second term of this Lagrangian contributes to the so-called $Y$ parameter
\cite{Barbieri:2004qk},
\begin{equation}
\begin{split}
\\[-2.5ex]
Y = \left( \frac{g'_Y}{g''_Y} \frac{m_W}{ m_{B^-}} \right)^2 \simeq \left( \frac{g'_Y}{g''_Y} \frac{m_W}{ m_{Z'}} \right)^2 \, , \\[1ex]
\end{split}
\label{yparameter}
\end{equation}
where $m_W$ is the $W$ boson mass.
The direct constraint on the $Y$ parameter is $Y = (4.2 \pm 4.9) \times 10^{-3}$.
Note that a similar analysis applies in any model with a $Z'$ which
mixes only through the $U(1)$.

\subsection{The $Z'$ mass bound\label{subsec:zprime}}

The $Z'$ boson is mainly produced by Drell-Yan like quark annihilation at the LHC.
This boson can decay into a pair of leptons.
The null result of dielectron and dimuon final state searches by the ATLAS and CMS detectors \cite{Dilepton,DileptonCMS}
gives the strongest bound on the $Z'$ mass.
From the interaction \eqref{Zprimeint}, the decay width of $Z'$ into a pair of fermions is given by
\begin{equation}
\begin{split}
\\[-2.5ex]
\Gamma (Z' \rightarrow f \bar{f} ) = \frac{C_f m_{Z'}}{24 \pi} \frac{(g'_Y)^4}{(g'_Y)^2 + (g''_Y)^2} \left( (Y^f_L)^2 + (Y^f_R)^2 \right)
\, , \\[1ex]
\end{split}
\label{zprimewidth}
\end{equation}
where $C_f$ is the color factor ($1$ for a color singlet and $3$ for a color triplet).
The $Z'$ boson can also decay into two bosons, 
$Z' \rightarrow W^+ W^-, Zh, ZG_S$, if kinematically allowed.
These are not dominant in the most of the parameter space and do not
dramatically affect the branching ratio into leptons.
The coupling (\ref{Zprimeint}) also implies the Drell-Yan production rate
of the $Z'$ is inversely proportional to $g_H^2$ for large $g_H$.  But this
does not help much.  If the $Z'$ decays dominantly into standard model
particles, the branching ratio into leptons is large and a $Z'$ lighter
than a few TeV is ruled out~\cite{Aad:2014fha, Aad:2014cka}.
However, if we introduce new fermions charged under the $SU(N+3)_{H}$ 
gauge group, as we do in section~\ref{sec:xdecay}, the coupling of the $Z'$
to these is proportional to $g_H$, and therefore much larger than the
coupling to standard model particles.  
If the $Z'$  decay into these fermions is kinematically allowed, it
dominates over the standard model decays in the interesting region of large
$g_H$ and a light $Z'$ is not impossible.


\subsection{Coloron phenomenology\label{subsec:coloron}}

Let us look at the color octet massive vector bosons, colorons, which are also mainly produced by quark annihilation at the LHC.
The NLO cross section of coloron production from quark annihilation has been calculated in Ref~\cite{Chivukula:2013xla}.
The gluon fusion contribution has been analyzed in Ref~\cite{Chivukula:2011ng}
and gives a sub-leading effect.
The coloron can decay into $G_O G_O$, $G_O Z'$, $q\bar{q}$ and $X'\bar{X}'$ if these decay modes are open.
The relevant interactions of these decay modes are summarized in appendix~\ref{coloronintsummary}.
The two-body decay rates of the coloron are then given by
\begin{equation}
\begin{split}
\\[-2.5ex]
&\Gamma (G' \rightarrow G_O G_O) = \frac{1}{256 \pi} \frac{\left( g_H^2 - (g'_3)^2 \right)^2}{g_H^2 + (g'_3)^2} \, m_{G'}
\left( 1 - \frac{4 m_{G_O}^2}{m_{G'}^2} \right)^{3/2} \, , \\[1.5ex]
&\Gamma (G' \rightarrow G_O Z') = \frac{1}{36 \pi} \left( g_H^2 + (g'_3)^2 \right) \frac{m_{Z'}^2}{m_{G'}^2}
\, {\rm \bf p} \left( 3 + \frac{{\rm \bf p}^2}{m_{Z'}^2} \right)  \, , \\[1.5ex]
&\Gamma (G' \rightarrow q \bar{q}) = \frac{1}{24 \pi} \frac{(g'_3)^4}{g_H^2 + (g'_3)^2}  \, m_{G'}
\left( 1 - \frac{4 m_{q}^2}{m_{G'}^2} \right)^{1/2} \, , \nonumber \\[1ex]
\end{split}
\end{equation}
\begin{equation}
\begin{split}
\\[-2.5ex]
&\Gamma (G' \rightarrow X' \bar{X}') = \frac{N}{96\pi} \frac{g_H^4}{g_H^2 + (g'_3)^2} \, m_{G'}
\left( 1- \frac{4 m_{X'}^2}{m_{G'}^2} \right)^{3/2} \left( 3 - \frac{m_{G'}^2}{m_{X'}^2} + \frac{m_{G'}^4}{4 m_{X'}^4} \right) \, , \\[1ex]
\end{split}
\end{equation}
where $m_q$ is the quark mass and
\begin{equation}
\begin{split}
\\[-2.5ex]
{\rm \bf p}^2 = \frac{1}{4 m_{G'}^2} \left( m_{G'}^2 - (m_{Z'} - m_{G_O})^2 \right)
\left( m_{G'}^2 - (m_{Z'} + m_{G_O})^2 \right) \, . \\[1ex]
\end{split}
\end{equation}
Because of (\ref{largeghmasses}), we do not expect the $G' \rightarrow X'
\bar{X}'$ to be allowed in the interesting region of large $g_H$.  
As in the case of the $Z'$ boson, 
if we introduce new fermions charged under the $SU(N+3)_{H}$ gauge group,
$G'$ can also decay into the quark components of the new fermions.  
Because the $m_{Z'}<m_{G'}$
for large $g_H$ (by (\ref{largeghmasses}) again), the coloron decay is
kinematically allowed whenever the $Z'$ decay is.
Thus if we introduce new $SU(N+3)$ fermions to evade the $Z'$ search
bounds, we will automaatically evade the coloron search bounds. 
If the $G_O$ is very light, the  $G' \rightarrow G_O G_O$ mode and perhaps
$G' \rightarrow G_O Z'$ can be important.

Another experimental constraint on the coloron mass and 
its interactions with the standard model quarks
comes from searches for quark contact interactions. 
The coloron exchange induces four-fermion interactions among the quarks,
\begin{equation}
\begin{split}
\\[-2.5ex]
\mathcal{L}_{\rm eff} \, \supset \, - \frac{(g'_3)^4}{{g_H^2 + (g'_3)^2}} \frac{1}{m_{G'}^2} 
\left( \bar{q} \gamma_\mu T^a q \right) \left( \bar{q} \gamma^\mu T^a q \right) \, . \\[1ex]
\end{split}
\label{coloroncontact}
\end{equation}
These quark contact interactions lead to constructive interference with the
ordinary QCD terms 
and hence deviation of dijet angular distributions from the perturbative
QCD predictions. 

There is certainly a strong constraint on (\ref{coloroncontact}) from LHC
data.  Unfortunately, the published results
from CMS in \cite{Khachatryan:2014cja}
consider only a set of contact terms which they call ``the most general
flavor diagonal'' set, but which is not general enough to include
(\ref{coloroncontact}).  This poor choice also appears in the particle data
group review of compositeness~\cite{Hagiwara:1998ka}. 
A sensible general form appears in
\cite{Eichten:1984eu}, but unfortunately this does not seem to have been
universally adopted in the literature.
We expect that the constraint on (\ref{coloroncontact}) will be of the same
order of magnitude of those quoted in \cite{Khachatryan:2014cja}.
\begin{equation}
a=\frac{m_{G'}}{\sqrt{g_H^2+(g'_3)^2}}
\, \gtrapprox \, \frac{(g'_3)^2}{g_H^2+(g'_3)^2}\,5\mbox{~TeV}
\label{coloroncontactbound}
\end{equation} 
This constraint is not affected (at least not very much) by the additional
$SU(N+3)$ fermions that we will introduce in section~\ref{sec:xdecay}.

Note that this constraint gives a very severe lower bound on the scale $a$ in
the small $g_H$ region of our parameter space 
because the coloron mass is approximately given by $m_{G'} \approx g'_3 a$
in this region. But for large $g_H$, relatively light colorons may be allowed.

\section{$N$-Glueballs and the 750 GeV diphoton excess\label{sec:glueballs}}

We here consider phenomenology of the glueballs associated with the
$SU(N)_H$ gauge theory, which we call $N$-glueballs, 
and their possible explanation of the $750 \, \rm GeV$ diphoton excess
observed at the LHC. 
First, we discuss the mass spectrum of the $N$-glueballs.
Then, we analyze the effective higher dimensional operators
involving $N$-gluons and the standard model particles 
which are relevant for the glueball decays.
We find a region of 
parameter space where the lightest glueball at around $750 \, \rm
GeV$ could explain the diphoton excess 
while satisfying the experimental constraints discussed above.
The decays of the pseudoscalar and spin $2$ glueballs are also presented.

\subsection{The $N$-glueball masses\label{subsec:glueballmass}}

Below the scale of the $SU(N+3)_H$ symmetry breaking, 
the unbroken $SU(N)_H$ gauge interaction becomes strong and finally
confines
giving rise to the $N$-glueball spectrum.
For very small $g_H$ coupling, the confinement scale of the $SU(N)_H$ pure
Yang-Mills gauge theory, 
denoted as $\Lambda_H$, is generically well below the symmetry breaking
scale, and we can estimate it using $0$-loop matching and the $1$-loop
$\beta$-function:  
\begin{equation}
\begin{split}
\\[-2.5ex]
\Lambda_H = m_{X'}\, e^{-\frac{6\pi}{(11N-2n_f) \alpha_H (a)}}  \, . \\[1ex]
\end{split}
\label{0loop}
\end{equation}
Here, $\alpha_H(a) \equiv g_H^2 (a) / 4 \pi$ means the gauge coupling at the scale $a$ and $n_f$  is
the number of $SU(N)$ fermions in the low-energy theory. 
Note that the confinement scale is scheme independent at 1-loop level.
We could improve on (\ref{0loop}) using the techniques of Hall and Weinberg
\cite{Hall:1980kf,Weinberg:1980wa} including $1$-loop matching and
$2$-loop renormalization, but this will not change the qualitative message
of (\ref{0loop}).  $\Lambda_H$ is smaller than $m_{X'}$, but for large
$\alpha_H$, we would expect the exponential factor in (\ref{0loop}) to be
of order 1 unless the running in the low-energy theory is very slow, for
example by having matter fields to nearly cancel the effect of $SU(N)_H$
gauge fields.

For a given $\Lambda_H$, we can appeal to lattice calculations to estimate
the glueball masses.
From 
\cite{Chen:2005mg}, the scalar glueball $0^{++}$ is the lightest and its
mass $m_0$ is estimated to lie in the region 
$4.7 \Lambda^{\overline{MS}} < m_0 < 11 \Lambda^{\overline{MS}}$
($\Lambda^{\overline{MS}}$ is the $\overline{MS}$ scheme confinement scale)
with very small dependence on $N$.
From the lattice result
\cite{Morningstar:1999rf},
the spin $2^{++}$ glueball mass is $m_{2^{++}} \simeq 1.4 \, m_0$
and the pseudoscalar glueball mass is $m_{0^{-+}} \simeq 1.5 \, m_0$.
There are many other states but we concentrate on these three lightest
$N$-glueballs in the rest of the discussion. 

As we have seen in the discussion of experimental constraints, and will
emphasize below, the interesting parameter space that might explain the
diphoton excess is in large $\alpha_H$ region.
In this region, our theory is strongly coupled and 
(\ref{0loop}) is certainly a reliable quantitative guide.
It is even unclear that the relevant symmetry breaking takes place as the
perturbation theory suggests. 
Thus, we do not know the relation
between the $X'$ mass and the glueball mass.
We will simply
assume that the $X'$ mass and the glueball mass are of the same order
and in the interesting region for the diphoton excess. 

\subsection{The dimension eight operators\label{subsec:dim8}}

\renewcommand{\arraystretch}{1.3}
\begin{table}[!t]
\begin{center}
\begin{tabular}{c|c}
$J^{PC}$ & Operator
 \\
 \hline
  $0^{++}$ & $S = \tr \, F_{\mu\nu} F^{\mu\nu}$  \\[1ex]
  $0^{-+}$ & $P = \tr \, F_{\mu\nu} \tilde{F}^{\mu\nu}$  \\[1ex]
  $2^{++}$, $1^{-+}$, $0^{++}$ & $T_{\mu \rho} = \tr \, F_{\mu \lambda} F_{\rho}^{\,\, \lambda} - \frac{1}{4} g_{\mu\rho} S$  \\[1ex]
  $2^{++}$, $2^{-+}$ & $L_{\mu\nu\rho\sigma} = \tr \, F_{\mu\nu} F_{\rho \sigma} - \frac{1}{2} \left( g_{\mu \rho}
T_{\nu \sigma} + g_{\nu \sigma} T_{\mu \rho} - g_{\mu \sigma} T_{\nu \rho} - g_{\nu \rho} T_{\mu \sigma} \right) $  \\
& $- \frac{1}{12} \left( g_{\mu \rho} g_{\nu \sigma} - g_{\mu \sigma} g_{\nu \rho} \right) S + \frac{1}{12} \epsilon_{\mu\nu\rho\sigma} P$
\end{tabular}
\end{center}
\caption{The dimension four operator which represents each glueball state.
Here, $F_{\mu \nu}$ denotes the field strength of the $SU(N)_H$ gauge boson
and $\tilde{F}_{\mu\nu} = \frac{1}{2} \epsilon_{\mu\nu\rho\sigma} F^{\rho \sigma}$.
The trace acts on the $SU(N)_H$ generators.}
\label{tab:glueoperator}
\end{table}
\renewcommand{\arraystretch}{1}

\renewcommand{\arraystretch}{1.3}
\begin{table}[!t]
\begin{center}
\vspace{0.3cm}
\begin{tabular}{c|c}
$J^{PC}$ & Operator
 \\
 \hline
 $1^{--}$, $1^{+-}$ & $\Omega^{(1)}_{\mu\nu} = \tr \, F_{\mu\nu} F_{\rho \sigma} F^{\rho \sigma} $  \\[1ex]
 $1^{--}$, $1^{+-}$ & $\Omega^{(2)}_{\mu\nu} = \tr \, F_{\mu}^{\,\, \rho} F_{\rho}^{\,\, \sigma} F_{\sigma \nu} $  \\
\end{tabular}
\end{center}
\caption{The dimension six operator which represents each glueball state.}
\label{tab:gluesix}
\end{table}
\renewcommand{\arraystretch}{1}

\begin{figure}[!t]
\vspace{-1cm}
\hspace{-0.5cm}
 \begin{minipage}{0.5\hsize}
  \begin{center}
   \includegraphics[clip, width=9cm]{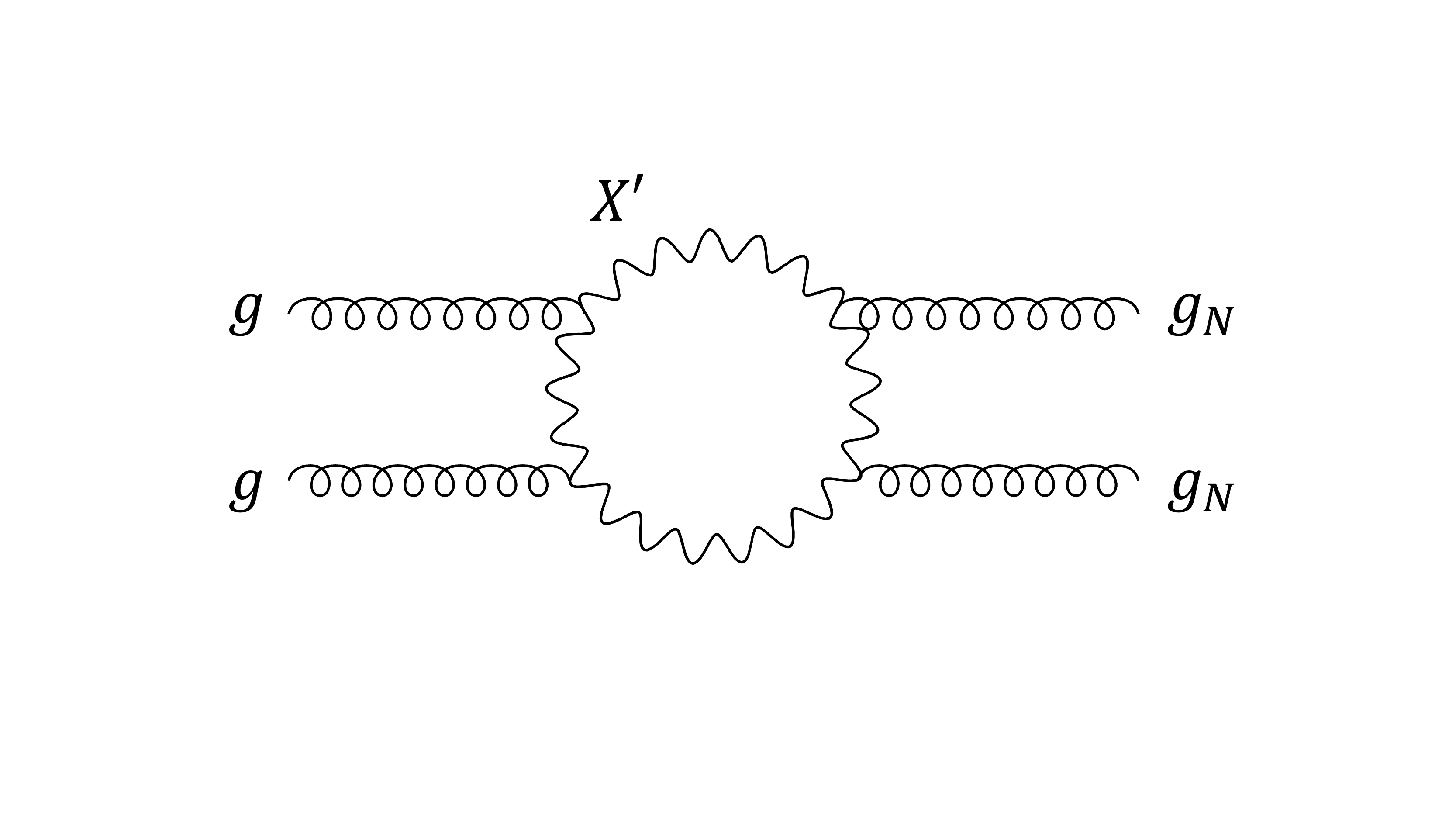}
  \end{center}
 \end{minipage}
 \begin{minipage}{0.5\hsize}
  \begin{center}
  {\vspace{-1mm}\includegraphics[clip, width=9cm]{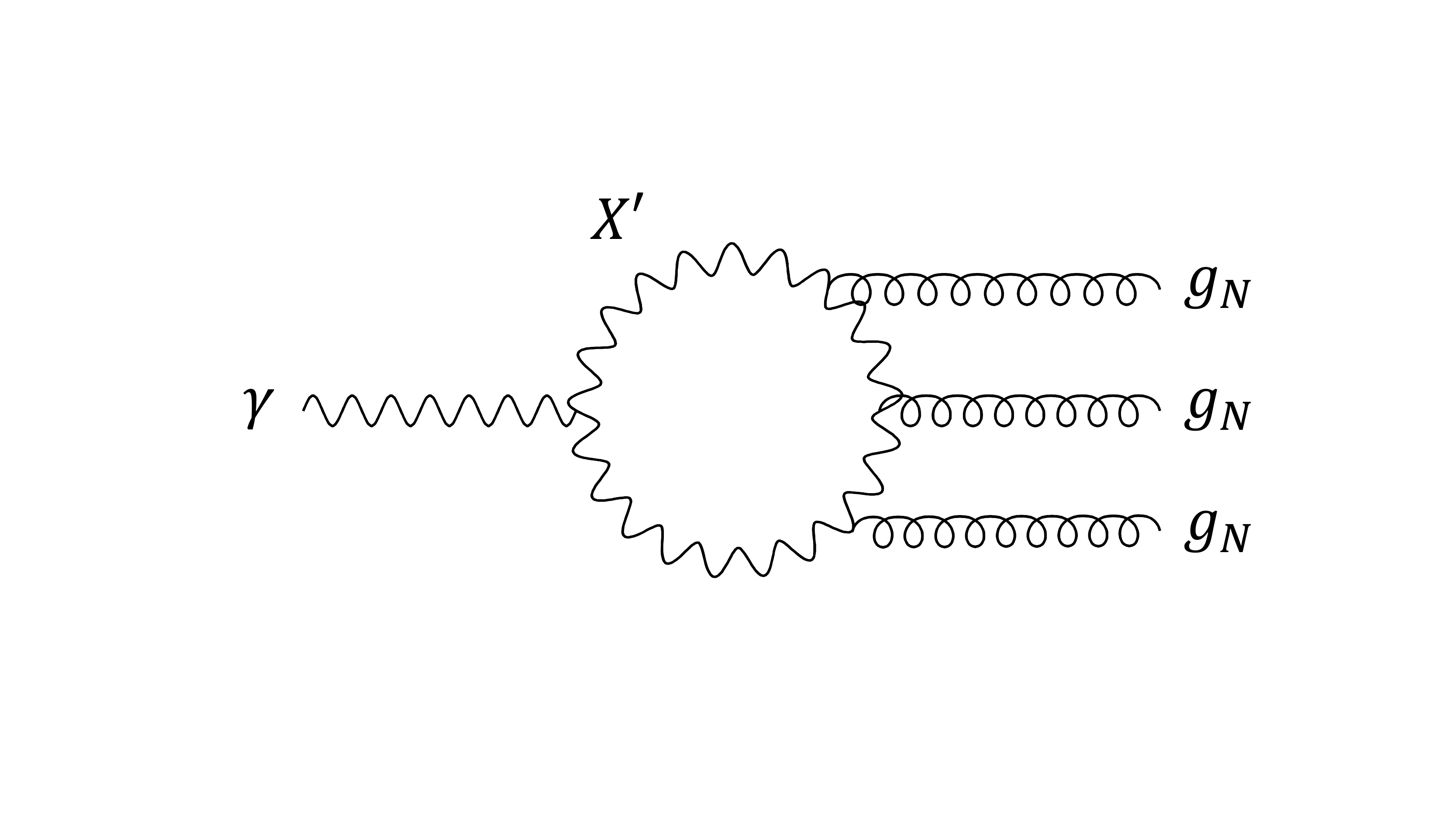}}
  \end{center}
 \end{minipage}
\vspace{-0.5cm}
  \caption{Two example diagrams of the $X', \bar{X}'$ gauge boson loops to generate
the effective dimension eight operators of the $N$-glueballs with the standard model gauge fields.
Here, $g$, $\gamma$ and $g_N$ denote the ordinary gluon, the photon and the $SU(N)_H$ gauge boson respectively.}
  \label{fig:XloopDimEight}
\vspace{1cm}
\end{figure}

In our model, the $N$-glueballs can decay into the standard model gauge bosons
through loops of the $X', \bar{X}'$ gauge bosons.
When the confinement scale $\Lambda_H$ is sufficiently small compared to the scale $a$,
the situation is similar to the so-called Hidden Valley scenario
\cite{Strassler:2006im} where the $X', \bar{X}'$ gauge bosons correspond to mediators
between the standard model sector and the hidden $SU(N)_H$ gauge sector.
Ref~\cite{Juknevich:2009ji} has discussed the hidden glueball decays into the standard model gauge bosons
through loops of heavy fermions.
In ref~\cite{Juknevich:2009ji},
these decays are analyzed using the factorized matrix elements,
\begin{equation}
\begin{split}
\\[-2.5ex]
&\mathcal{M} (\Psi \rightarrow \mathcal{A} \mathcal{A}) =
\langle {SM} | \mathcal{O}_{SM} | 0 \rangle \langle 0 | \mathcal{O}_H | \Psi \rangle \, , \\[1ex]
&\mathcal{M} (\Psi \rightarrow \Psi' + \mathcal{A}) =
\langle {SM} | \mathcal{O}_{SM} | 0 \rangle \langle \Psi' | \mathcal{O}_H | \Psi \rangle \, . \label{factorized} \\[1ex]
\end{split}
\end{equation}
Here, $\Psi^{(')}$ denotes a glueball state and $\mathcal{A}$ the standard model gauge bosons collectively.
After integrating out heavy fields in the loops, the decays are described by dimension eight operators,
$\mathcal{L}_{\rm eff} \supset \mathcal{O}_{SM} \mathcal{O}_H$
where $\mathcal{O}_{SM}$ represents an operator composed of the standard model gauge fields.
Table~\ref{tab:glueoperator} (\ref{tab:gluesix}) shows the relevant dimension four (six) operator $\mathcal{O}_H$
which represents each glueball state
\cite{Jaffe:1985qp,Juknevich:2009ji}.
Then, the effective Lagrangian after integrating out the $X', \bar{X}'$ gauge bosons can be written as
\begin{equation}
\begin{split}
\\[-2.5ex]
\mathcal{L}_{\rm eff} = \, &\frac{g_H^2}{(4\pi)^2 m_{X'}^4}
\left( g_Y^2 \kappa_Y B^{\mu\nu} B^{\rho\sigma}
+ g_s^2 \kappa_s \tr \,  G^{\mu\nu} G^{\rho\sigma}  \right) \\[1ex]
&\times \left( a_S S g_{\mu\rho} g_{\nu\sigma} + a_P P \epsilon_{\mu\nu\rho\sigma} 
+ a_T T_{\mu\rho} g_{\nu\sigma} + a_L L_{\mu\nu\rho\sigma} \right) \\[1ex]
&+ \frac{g_H^3 g_Y}{(4\pi)^2 m_{X'}^4} \kappa_\Omega \left( b_1 B^{\mu\nu} \Omega^{(1)}_{\mu\nu}
+ b_2 B^{\mu\nu} \Omega^{(2)}_{\mu\nu}  \right)  \, , \label{dimeight} \\[1ex]
\end{split}
\end{equation}
where $B_{\mu\nu}$ and $G_{\mu\nu}$ denote the field strengths of the ordinary hypercharge and color gauge fields
and $\kappa_Y = 6 q^2$, $\kappa_s = 2$ and $\kappa_\Omega = 6 q$.
The coefficients $a_{S, P, T, L}$, $b_{1,2}$ are obtained by the one-loop computation.
Two examples of the relevant diagrams of the $X', \bar{X}'$ gauge boson loops are shown in Figure~\ref{fig:XloopDimEight}.
The calculation of the coefficients has been done in
\cite{Metsaev:1987ju,Fichet:2013ola} 
and is summarized in appendix~\ref{sec:dimeight}.
The results are
\begin{equation}
\begin{split}
\\[-2.5ex]
&a_S = \frac{89}{480} \, , \qquad  a_P = \frac{79}{960} \, , \qquad  a_T = \frac{7}{5} \, , \qquad  a_L = \frac{1}{40} \, , \\[2ex]
&b_1 = -\frac{5}{16} \, , \qquad  b_2 = \frac{27}{20}
\, . \label{coefficients} \\[1ex]
\end{split}
\end{equation}
The coefficient $a_S$ here is about a factor of ten larger than when
particles inside the loops are fermions 
($a_S |_{\rm fermion} = 1/60$ \cite{Juknevich:2009ji}).
Thus the production cross section of the lightest glueball by gluon fusion
is enhanced by a factor of $\mathcal{O}(100)$. 
This is one of the promising features in this model for the explanation of
the reported diphoton excess. 

\subsection{The scalar effective operator\label{subsec:eft}}

\begin{figure}[!t]
\vspace{-1cm}
 \begin{minipage}{0.5\hsize}
  \begin{center}
   \includegraphics[clip, width=9cm]{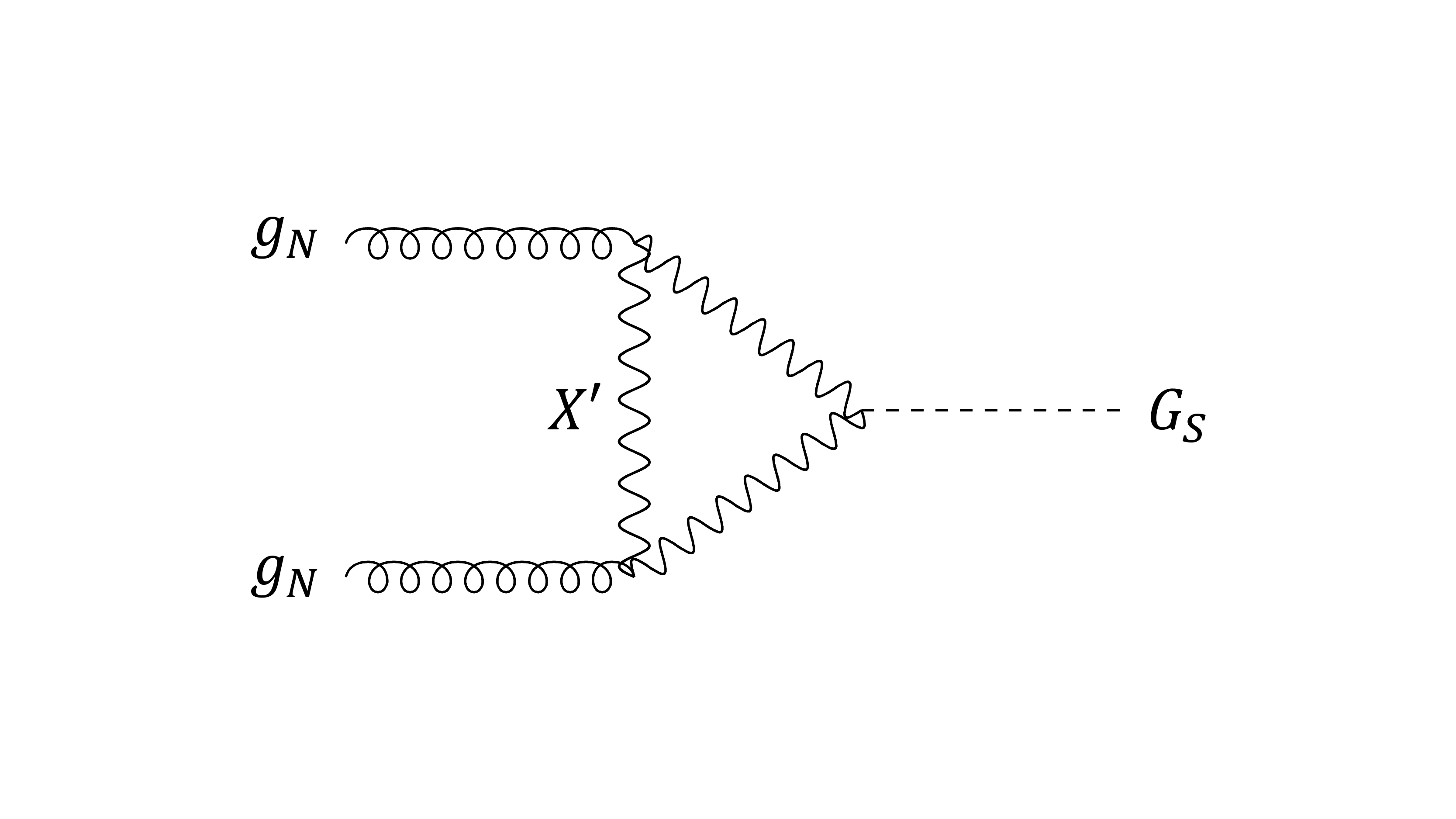}
  \end{center}
 \end{minipage}
 \begin{minipage}{0.5\hsize}
  \begin{center}
  \includegraphics[clip, width=9cm]{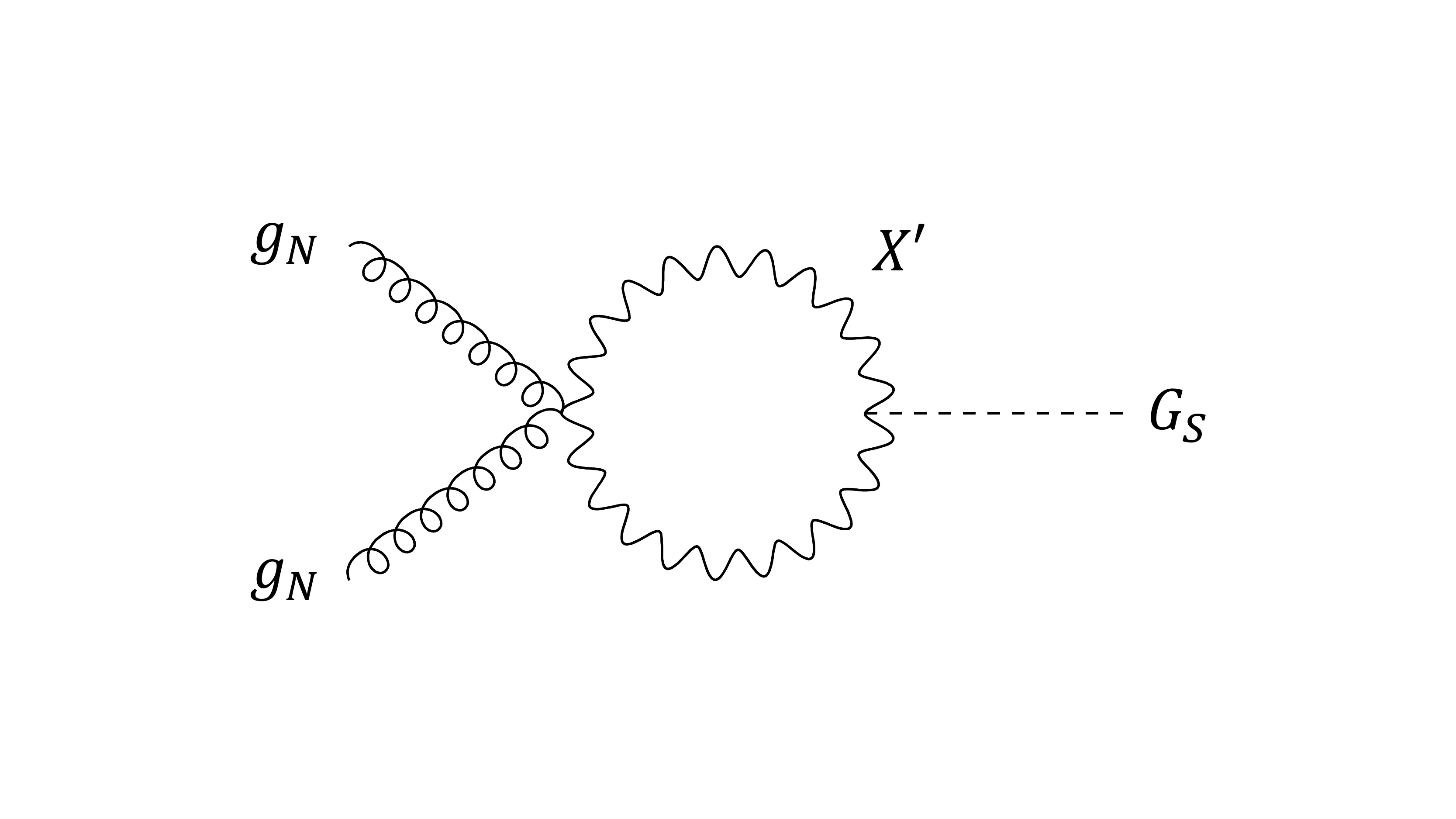}
  \end{center}
 \end{minipage}
\vspace{-0.5cm}
  \caption{One-loop diagrams of the $X'$ gauge boson to contribute
to the effective interaction of the $SU(N)_H$ gauge fields with the color singlet scalar $G_S$.
There are also the diagrams of the $\bar{X}'$ gauge boson.}
  \label{fig:Xloop}
\vspace{1cm}
\end{figure}

The mixing between the scalar $N$-glueball and the singlet scalar $G_S$ is generated by loops of the $X', \bar{X}'$ gauge bosons.
This may be important for the glueball decays because the singlet $G_S$ also mixes with the Higgs boson
and the glueball decays into a pair of the standard model fermions and massive gauge bosons are induced through these mixings.
The one-loop diagrams of the massive vector boson $X'$ to generate the effective interaction of the $SU(N)_H$ gauge fields with $G_S$
are shown in Figure~\ref{fig:Xloop} (There are also the diagrams of the $\bar{X}'$ gauge boson).
The relevant interactions of the $X', \bar{X}'$ gauge bosons with the scalar $G_S$ and the $SU(N)_H$ gauge fields are summarized in
appendix~\ref{XXScalar}.
The similar calculation as the case of the Higgs boson decays through the $W$ boson loops gives
the mixing term between $G_S$ and the scalar glueball,
\begin{equation}
\begin{split}
\\[-2.5ex]
\mathcal{L}_{G_S-S} \, = \, \frac{\alpha_H}{2 \pi}\frac{k_{g_N}}{\Lambda_{g_N}} \, G_S \, S \, , \quad \qquad 
\frac{k_{g_N}}{\Lambda_{g_N}}  = - \left( \frac{3}{4\sqrt{6}  a} \right) F_{V} (\tau_{X'}) \, . \label{glueballmixing} \\[1ex]
\end{split}
\end{equation}
Here, we have defined $\tau_{X'} \equiv m_{G_S}^2 / (4 m_{X'}^2)$.
The loop function $F_{V} (\tau)$ is given by
\begin{equation}
\begin{split}
\\[-2.5ex]
F_{V} (\tau) = - \left( \tau^{-1} (3 + 2\tau) + 3 \tau^{-2} (-1 + 2 \tau) Z(\tau) \right) \, , \\[1ex]
\end{split}
\end{equation}
and 
\begin{equation}
\begin{split}
\\[-2.5ex]
Z (\tau) =
  \begin{cases}
    \left[ \sin^{-1} (\sqrt{\tau})   \right]^2  &(\tau \leq 1) \\[1ex]
   - \frac{1}{4} \log \left[ \frac{1 + \sqrt{1 - \tau^{-1}}}{1 - \sqrt{1 - \tau^{-1}}} - i \pi\right]^2 &(\tau \geq 1)
  \end{cases}
\, . \\[1ex]
\end{split}
\end{equation}
Using this effective interaction, we will discuss the lightest $N$-glueball decays
into a pair of the standard model fermions and massive gauge bosons.

We here comment on phenomenology of the singlet scalar $G_S$ briefly.
The $G_S$ scalar can be produced by gluon fusion through loops of both the coloron and the $X', \bar{X}'$ gauge bosons.
The produced $G_S$ decays into a pair of the standard model gauge bosons and the Higgs bosons.
The decays into the standard model fermions and massive gauge bosons are also possible through the mixing with the Higgs boson.
Furthermore, when the mass of $G_S$ is larger than twice the $N$-glueball mass,
the same loops of the $X', \bar{X}'$ gauge bosons as above induce the $G_S$ decay into two glueballs.
We leave detailed phenomenology of the scalar $G_S$ to a future study.

\subsection{The lightest glueball decays\label{subsubsec:lightest}}

We now consider the decays of the lightest $N$-glueball $0^{++}$
through the effective dimension eight operators in \eqref{dimeight} generated by loops of the $X', \bar{X}'$ vector bosons.
The glueball dominantly decays into a pair of gluons.
The diphoton decay is also induced by the new vector boson loops.
As discussed above, the decay amplitude is written by the factorized matrix element \eqref{factorized}.
The amplitude of the glueball decay into a pair of gluons is then given by
\begin{equation}
\begin{split}
\\[-2.5ex]
\mathcal{M} (0^{++} \rightarrow gg) = \frac{\alpha_s \alpha_H}{m_{X'}^4} \kappa_s a_S \,
\langle g_1^a g_2^b | \tr \, G_{\mu\nu} G^{\mu\nu} | 0 \rangle \langle 0 | S | 0^{++} \rangle
\, . \\[1ex]
\end{split}
\end{equation}
Here, the transition to two gluons is
$\langle g_1^a g_2^b | \tr \, G_{\mu\nu} G^{\mu\nu} | 0 \rangle = \delta^{ab} \left( k^1_\mu \epsilon^1_\nu
- k^1_\nu \epsilon^1_\mu \right) \left( k^{2\mu} \epsilon^{2\nu} - k^{2\nu} \epsilon^{2\mu} \right)$
where $k^{1,2}$ are gluon momenta and $\epsilon^{1,2}$ are polarizations.
From this decay amplitude, the decay rate is calculated as
\begin{equation}
\begin{split}
\\[-2.5ex]
\Gamma (0^{++} \rightarrow gg) = \frac{8 \alpha_s^2 \alpha_H^2}{16\pi m_{X'}^8} \kappa_s^2 a_S^2 m_0^3 \left({\bf F^S_{0^{++}}} \right)^2
\, , \label{gluegg} \\[1ex]
\end{split}
\end{equation}
where $ {\bf F^S_{0^{++}}} \equiv \langle 0 | S | 0^{++} \rangle$ is the decay constant of the scalar glueball $0^{++}$
and $2.00 \, m_0^3 \leq 4 \pi \alpha_H {\bf F^S_{0^{++}}} \leq 4.77 \, m_0^3$ from the lattice result for the $SU(3)$ pure Yang-Mills theory
\cite{Chen:2005mg}.
We assume that this lattice result persists also in cases with general numbers of $N$.
In the same way, we can compute the decay rates of $0^{++} \rightarrow \gamma \gamma , ZZ , Z \gamma$.
The branching ratios are given by
\begin{equation}
\begin{split}
&{\rm Br}(0^{++} \rightarrow \gamma \gamma) \simeq
\frac{\Gamma (0^{++} \rightarrow \gamma \gamma)}{\Gamma (0^{++} \rightarrow gg)} 
= \frac{\alpha^2}{2\alpha_s^2} \frac{\kappa_{Y}^2}{\kappa_s^2} = \frac{9}{2} \frac{q^4\alpha^2}{\alpha_s^2}
\, , \label{gluegammagamma} \\[1ex]
\end{split}
\end{equation}
and
\begin{equation}
\begin{split}
\\[-2.5ex]
&{\rm Br}(0^{++} \rightarrow ZZ) \simeq
\frac{\Gamma (0^{++} \rightarrow ZZ)}{\Gamma (0^{++} \rightarrow gg)} 
= \frac{\alpha^2 \tan^4 \theta_W }{2 \alpha_s^2} \frac{\kappa_Y^2}{\kappa_s^2} \left( 1 - \frac{4 m_Z^2}{m_0^2} \right)^{1/2}
\left( 1 - \frac{4 m_Z^2}{m_0^2} + \frac{6 m_Z^4}{m_0^4} \right)
\, , \\[2ex]
&{\rm Br}(0^{++} \rightarrow Z\gamma) \simeq
\frac{\Gamma (0^{++} \rightarrow Z\gamma)}{\Gamma (0^{++} \rightarrow gg)} 
= \frac{\alpha^2 \tan^2 \theta_W }{\alpha_s^2} \frac{\kappa_Y^2}{\kappa_s^2}  \left( 1 - \frac{m_Z^2}{m_0^2} \right)^{3}
\, . \\[1ex]
\end{split}
\end{equation}
Here, we have assumed that the total decay width is approximately given by $\Gamma_{\rm total} \simeq \Gamma (0^{++} \rightarrow gg)$.
These decay modes are also induced through the glueball mixing with the $G_S$ scalar but
they are effectively two-loop effects and can be ignored.
Note that the branching ratio of the diphoton decay is completely determined by the electric charge $q$ of the $X'$ gauge boson
unlike the case where particles in the loops are various fermions with various masses and charges.

From the mixing term \eqref{glueballmixing} generated by loops of the $X',
\bar{X}'$ gauge bosons, 
the glueball decays $0^{++} \rightarrow hh, f \bar{f} , WW$ are also possible.
The decays $G_S \rightarrow hh, f \bar{f} , WW$ are induced by the $G_S$
interaction and mixing with the Higgs boson.  All of these depend on the
coupling $\lambda_4$ that governs $G_S$-$h$ mixing, so they need not be large. 
At leading order in $\lambda_4$, the decay rates of $0^{++} \rightarrow hh, f
\bar{f} , WW$ are written as 
\begin{equation}
\begin{split}
\\[-2.5ex]
\Gamma (0^{++} \rightarrow hh) &= \left( \frac{2 \alpha_H k_{g_N} {\bf F^S_{0^{++}}}}{4 \pi \Lambda_{g_N}(m_{G_S}^2 - m_0^2)} \right)^2
\Gamma_{G_S \rightarrow hh} (m_0^2)  \\[2ex]
&\simeq \left( \frac{2 \alpha_H k_{g_N} {\bf F^S_{0^{++}}}}{4 \pi \Lambda_{g_N}(m_{G_S}^2 - m_0^2)} \right)^2
\frac{\left( \sqrt{6} \lambda_4 a /2\right)^2}{32 \pi  m_{0}}
\sqrt{1 - \frac{4 m_{h}^2}{m_{0}^2} }
\, , \\[2ex]
\Gamma (0^{++} \rightarrow f \bar{f}) &= \left( \frac{2 \alpha_H k_{g_N} {\bf F^S_{0^{++}}}}{4 \pi \Lambda_{g_N}(m_{G_S}^2 - m_0^2)} \right)^2
\Gamma_{G_S \rightarrow f \bar{f}} (m_0^2) \\[2ex]
&\simeq \left( \frac{2 \alpha_H k_{g_N} {\bf F^S_{0^{++}}}}{4 \pi \Lambda_{g_N}(m_{G_S}^2 - m_0^2)} \right)^2
\left( \frac{3 \sqrt{6} \, \lambda_4}{9 \lambda_1 + 2 \lambda_2} \, \frac{v}{a}\right)^2  
\Gamma^{\rm SM}_{h \rightarrow f \bar{f}} (m_{0}^2)
\, , \\[2ex]
\Gamma (0^{++} \rightarrow WW ) &= \left( \frac{2 \alpha_H k_{g_N} {\bf F^S_{0^{++}}}}{4 \pi \Lambda_{g_N}(m_{G_S}^2 - m_0^2)} \right)^2
\Gamma_{G_S \rightarrow WW } (m_0^2)  \\[2ex]
&\simeq \left( \frac{2 \alpha_H k_{g_N} {\bf F^S_{0^{++}}}}{4 \pi \Lambda_{g_N}(m_{G_S}^2 - m_0^2)} \right)^2
\left( \frac{3 \sqrt{6} \, \lambda_4}{9 \lambda_1 + 2 \lambda_2} \, \frac{v}{a}\right)^2 
\Gamma^{\rm SM}_{h \rightarrow WW} (m_{0}^2)  \, . \\[1ex]
\end{split}
\end{equation}
Here, $\Gamma_{h \rightarrow f\bar{f}} (m_{0}^2)$ and $\Gamma_{h \rightarrow WW} (m_{0}^2)$ are the decay rates of
the Higgs boson into a pair of the standard model fermions and the $W$ bosons evaluated at the mass scale of the glueball.
The branching ratios of these decay modes depend on the parameters of the scalar sector.
In the rest of the discussion,
we assume the $\lambda_4$ coupling is not too large (or the $G_S$ scalar is heavy) so that they do not dominate over the diphoton decay.

The present calculations of the glueball decay rates only take into account the leading order effects.
At the next-to-leading order, we have substantial $\alpha_s$ and $\alpha_H$ corrections.
Then, the actual total decay rate of the lightest $N$-glueball may be larger.

\subsection{The diphoton excess\label{subsubsec:lightest2}}

\begin{figure}[!t]
  \begin{center}
  \vspace{0cm}
          \includegraphics[clip, width=9cm]{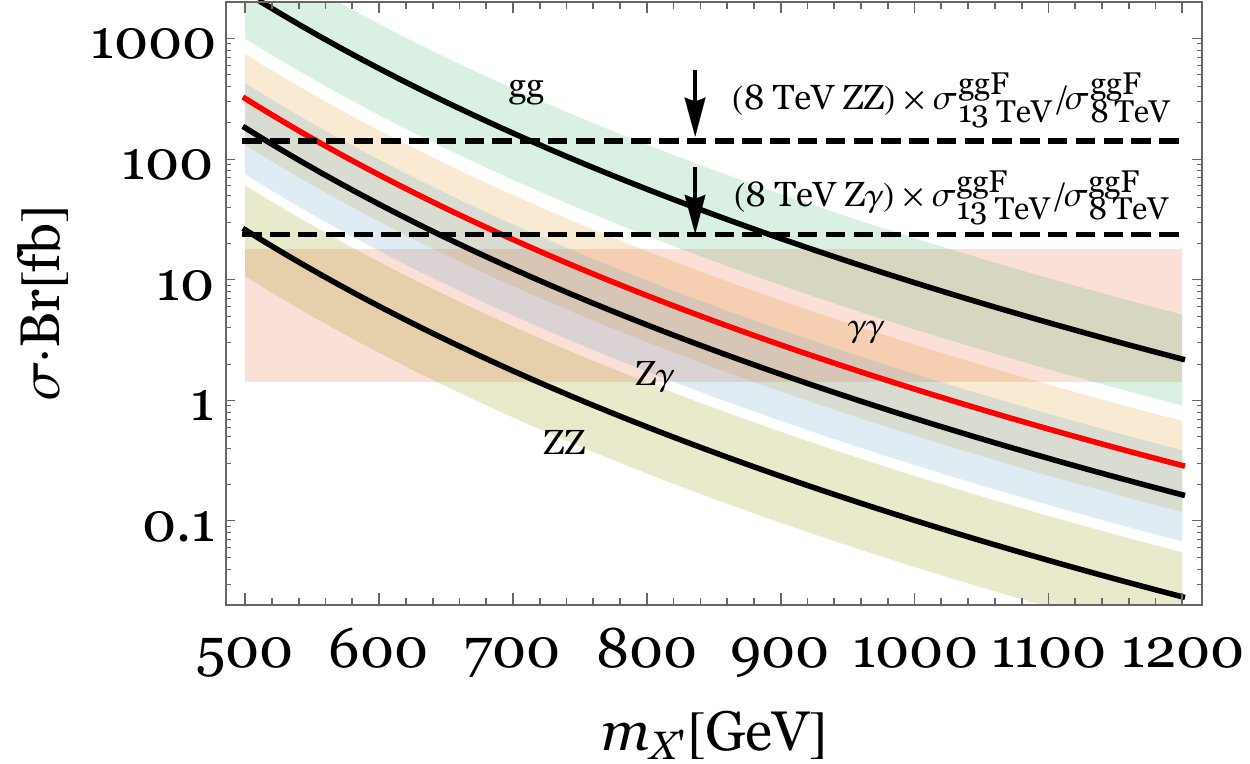}
    \caption{The production cross section times branching ratio
$\sigma(pp \rightarrow 0^{++}) \, {\rm Br} (0^{++} \rightarrow \gamma \gamma)$
at the $\sqrt{s} = 13 \, \rm TeV$ LHC.
We take the electric charge of the $X'$ gauge boson as $q=5/3$.
The resonance searches in $ZZ$ \cite{Aad:2015kna} and $Z(\rightarrow \ell \ell) \gamma$ \cite{Aad:2014fha} at $\sqrt{s} = 8 \, \rm TeV$
rescaled by the ratio $\sigma^{\rm ggF}_{13 \, \rm TeV}/\sigma^{\rm ggF}_{8 \, \rm TeV} \simeq 4.7$
put upper bounds on the production cross section times branching ratios.
Uncertainty of the glueball decay constant $ {\bf F^S_{0^{++}}}$ is included in each line
(which corresponds to $4 \pi \alpha_H {\bf F^S_{0^{++}}} = 3.12 \, m_0^3$).
The observed diphoton excess can be explained with $640 \, {\rm GeV} \lesssim m_{X'}
\lesssim 1100 \, {\rm GeV}$.}
    \label{fig:glueball}
  \end{center}
\end{figure}

\begin{figure}[!t]
  \begin{center}
  \vspace{0cm}
{ \includegraphics[clip, width=8cm]{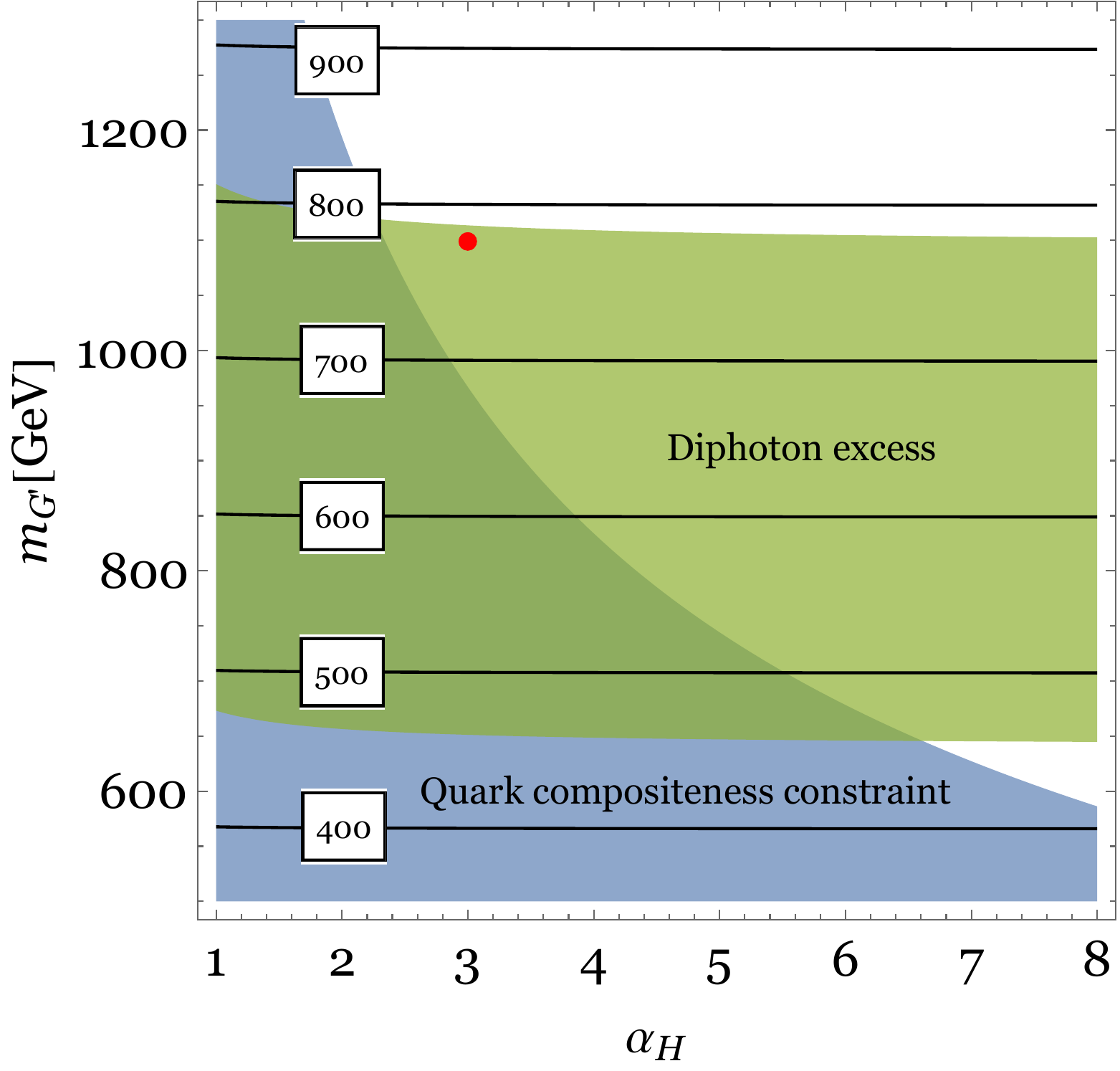}}
  \vspace{0cm}
    \caption{The region of $\alpha_H$
and the coloron mass $m_{G'}$
which can explain the reported diphoton excess (the green band).
Here, we assume $N=3$ and $q = 5/3$.
The blue shaded region is not allowed from the constraint on
the quark compositeness constraint.
The $Y$ parameter constraint is weak in this parameter region.
The model-dependent direct bounds on the $Z'$ mass and the coloron mass are
not ahown.
Although we do not know the precise mass relation between the coloron and the
glueballs for large $\alpha_H$, 
we assume that the lightest glueball mass is at $750 \, \rm GeV$ in the region
which can explain the diphoton excess.
The red dot denotes a benchmark point mentioned in the main text.}
    \label{fig:parameterspace}
  \end{center}
\end{figure}

Now we can put everything together and
discuss the possibility that the lightest $N$-glueball $0^{++}$ explains
an event excess in diphoton invariant mass distribution of around $750 \, \rm GeV$
reported by the ATLAS and CMS collaborations \cite{ATLAS,CMS:2016owr}.
The glueball can be produced by gluon fusion
through the effective dimension eight operator in \eqref{dimeight} generated by loops of the $X', \bar{X}'$ vector bosons.
With the narrow width approximation
\cite{Georgi:1977gs,Cahn:1983ip},
the production cross section times branching ratio $\sigma(pp \rightarrow 0^{++}) \, {\rm Br} (0^{++} \rightarrow \gamma \gamma)$
is expressed as
\begin{equation}
\begin{split}
\\[-2.5ex]
\sigma(pp \rightarrow 0^{++}) \, {\rm Br} (0^{++} \rightarrow \gamma \gamma)
&\simeq  \frac{\pi^2}{8 m_0 s} \, \Gamma (0^{++} \rightarrow \gamma \gamma) \\[1ex]
&\times \int^1_0 dx_1 \int^1_0 dx_2
 \left[ \delta(x_1 x_2 - m_0^2 / s) g (x_1) g(x_2) \right] \, , \\[1ex]
\end{split}
\end{equation}
where $s$ is the square of center of mass energy and
$g(x)$ is the parton distribution function (PDF) of the gluon.
We have assumed $\Gamma_{\rm total} \simeq \Gamma (0^{++} \rightarrow gg)$.
From \eqref{gluegg} and \eqref{gluegammagamma}, we can calculate
$\sigma(pp \rightarrow 0^{++}) \, {\rm Br} (0^{++} \rightarrow \gamma \gamma)$ with $m_0 = 750 \, \rm GeV$.
By using MSTW PDF \cite{Martin:2009iq}, this is numerically given by
\begin{equation}
\begin{split}
\\[-2.5ex]
\sigma(pp \rightarrow 0^{++}) \, {\rm Br} (0^{++} \rightarrow \gamma \gamma)
\, \simeq \,  2.9 \, {\rm fb} \times  \left( \frac{q}{5/3} \right)^4 \left( \frac{m_{X'}}{900 \, \rm GeV} \right)^{-8}
\quad (\sqrt{s} = 13 \, \rm TeV) \, . \\[1ex]
\end{split}
\end{equation}
The reported excess at $\sqrt{s} = 13 \, \rm TeV$ is \cite{Nakai:2015ptz}
\begin{equation}
\begin{split}
\\[-2.5ex]
1.4 \, {\rm fb} \, \lesssim \, \sigma(pp \rightarrow 0^{++}) \, {\rm Br} (0^{++} \rightarrow \gamma \gamma)
\, \lesssim \, 18 \, {\rm fb}  \, . \\[1ex]
\end{split}
\end{equation}
Figure~\ref{fig:glueball} shows the production cross section times branching ratio
$\sigma(pp \rightarrow 0^{++}) \, {\rm Br} (0^{++} \rightarrow \gamma \gamma)$
at the $\sqrt{s} = 13 \, \rm TeV$ LHC.
The shaded region denotes the observed excess.
We take the electric charge of the $X'$ gauge boson as $q=5/3$.
The resonance searches in $ZZ$ \cite{Aad:2015kna} and $Z(\rightarrow \ell \ell) \gamma$ \cite{Aad:2014fha} at $\sqrt{s} = 8 \, \rm TeV$
rescaled by the ratio $\sigma^{\rm ggF}_{13 \, \rm TeV}/\sigma^{\rm ggF}_{8 \, \rm TeV} \simeq 4.7$
put upper bounds on the production cross section times branching ratios.
Uncertainty of the glueball decay constant $ {\bf F^S_{0^{++}}}$ is included in each line
(which corresponds to $4 \pi \alpha_H {\bf F^S_{0^{++}}} = 3.12 \, m_0^3$).
The observed diphoton excess can be explained with $640 \, {\rm GeV} \lesssim m_{X'}
\lesssim 1100 \, {\rm GeV}$.
As the charge $q$ is large, the upper bound on the $X'$ mass is relaxed
as far as the upper bound on $q$ from \eqref{gYprime} is satisfied.

We now look at the model parameter space where the experimental constraints
are satisfied and 
the diphoton excess can be explained.
Figure~\ref{fig:parameterspace} shows the region of $\alpha_H$
and the coloron mass $m_{G'}$
which can explain the reported diphoton excess (the green band).
Here, we assume $N=3$ and $q = 5/3$.
The blue shaded region is not allowed from the constraint on
the quark contact interaction induced from the coloron exchange
\eqref{coloroncontactbound}. 
We can see that, as expected, this constraint pushes the interesting
parameter space to the large $\alpha_H$ region. 
The $Y$ parameter constraint is weak in this region.
The model-dependent search bounds on the $Z'$ mass and the coloron mass are not
shown,  
but are weak if the $Z'$ boson and the coloron decay into new fermions.
Although we do not know the precise relation between the coloron mass and the
glueball mass for large $\alpha_H$, 
we assume that the lightest glueball mass is at $750 \, \rm GeV$ in the
interesting parameter region that can explain the diphoton excess.

An interesting benchmark point, shown in figure~\ref{fig:parameterspace} (red dot),
is $\alpha_H=3$ and $m_{G'}=1100$~GeV.  This gives $m_{Z'}\approx777$~GeV,
$m_{X'}\approx1083$~GeV.  For this choice of parameter, $SU(N+3)$ fermions 
with mass between $375$~GeV and $m_{Z'}/2$ can dominate the $Z'$ and coloron
decay widths.   

\subsection{The pseudoscalar glueball decays\label{subsubsec:pseudo}}

We next consider the decays of the pseudoscalar $N$-glueball $0^{-+}$
through the effective dimension eight operator \eqref{dimeight}.
As in the case of the lightest scalar glueball, 
the width of the pseudoscalar glueball decay into a pair of gluons is 
\begin{equation}
\begin{split}
\\[-2.5ex]
\Gamma (0^{-+} \rightarrow gg) = \frac{8 \alpha_s^2 \alpha_H^2}{16\pi m_{X'}^8} \kappa_s^2 a_P^2 m_{0^{-+}}^3 \left({\bf F^P_{0^{-+}}} \right)^2
\, , \\[1ex]
\end{split}
\end{equation}
where ${\bf F^P_{0^{-+}}} \equiv \langle 0 | P | 0^{-+} \rangle$ is the decay constant of the pseudoscalar glueball.
We can also compute the decay rates of $0^{-+} \rightarrow \gamma \gamma , ZZ , Z \gamma$.
The branching ratios are given by
\begin{equation}
\begin{split}
\\[-2.5ex]
&{\rm Br}(0^{-+} \rightarrow \gamma \gamma) \simeq
\frac{\Gamma (0^{-+} \rightarrow \gamma \gamma)}{\Gamma (0^{-+} \rightarrow gg)} 
= \frac{\alpha^2}{2\alpha_s^2} \frac{\kappa_{Y}^2}{\kappa_s^2} = \frac{9}{2} \frac{q^4\alpha^2}{\alpha_s^2}
\, , \\[2ex]
&{\rm Br}(0^{-+} \rightarrow ZZ) \simeq
\frac{\Gamma (0^{-+} \rightarrow ZZ)}{\Gamma (0^{-+} \rightarrow gg)} 
= \frac{\alpha^2 \tan^4 \theta_W }{2 \alpha_s^2} \frac{\kappa_Y^2}{\kappa_s^2} \left( 1 - \frac{4 m_Z^2}{m_{0^{-+}}^2} \right)^{3/2}
\, , \nonumber \\[1ex]
\end{split}
\end{equation}
\begin{equation}
\begin{split}
\\[-2.5ex]
&{\rm Br}(0^{-+} \rightarrow Z\gamma) \simeq
\frac{\Gamma (0^{-+} \rightarrow Z\gamma)}{\Gamma (0^{-+} \rightarrow gg)} 
= \frac{\alpha^2 \tan^2 \theta_W }{\alpha_s^2} \frac{\kappa_Y^2}{\kappa_s^2}  \left( 1 - \frac{m_Z^2}{m_{0^{-+}}^2} \right)^{3}
\, . \\[1ex]
\end{split}
\end{equation}
The pseudoscalar glueball can also decay into the lightest glueball with a pair of gauge bosons,
but its branching ratio is significantly suppressed, as discussed in Ref~\cite{Juknevich:2009ji}.

When we fix the scalar glueball mass at $m_0 = 750 \, \rm GeV$,
the mass of the pseudoscalar glueball is $m_{0^{-+}} \simeq 1.1 \, \rm TeV $.
Numerically, the ratios are then calculated as
\begin{equation}
\begin{split}
\\[-2.5ex]
&{\rm Br}(0^{-+} \rightarrow gg) : {\rm Br}(0^{-+} \rightarrow \gamma \gamma) :
{\rm Br}(0^{-+} \rightarrow ZZ) : {\rm Br}(0^{-+} \rightarrow Z \gamma) \\[1ex]
&\simeq 1 - 0.028 \, q^4 : 0.017 \, q^4 : 0.0015 \, q^4 : 0.010 \, q^4 \, . \\[1ex]
\end{split}
\end{equation}
The branching ratios of $0^{-+} \rightarrow \gamma \gamma, Z \gamma$ are the same with those of the scalar glueball decays.
The decay to two gluons is dominant and the diphoton decay is the next.
The existence of this particle is one of the predictions in the glueball scenario of the reported diphoton excess.

\subsection{The $2^{++}$ glueball decays\label{subsubsec:spin2}}

Finally, we summarize the decays of the $2^{++}$ $N$-glueball.
The existence of this glueball is also a prediction in the present scenario.
The decay rates of $2^{++} \rightarrow gg, \gamma \gamma , ZZ , Z \gamma$ are calculated in Ref~\cite{Juknevich:2009ji}
for the case where particles inside the loops are fermions.
They are expressed in terms of the decay constants of the $2^{++}$ glueball,
\begin{equation}
\begin{split}
\\[-2.5ex]
&\langle 0 | T_{\mu\nu} | 2^{++} \rangle \equiv {\bf F^T_{2^{++}}} \epsilon_{\mu\nu} \, , \\[1ex]
&\langle 0 | L_{\mu\nu\rho \sigma} | 2^{++} \rangle \equiv {\bf F^L_{2^{++}}}
\left( \mathcal{P}_{\mu\rho} \epsilon_{\nu\sigma} - \mathcal{P}_{\mu\sigma} \epsilon_{\nu\rho} 
+ \mathcal{P}_{\nu\sigma} \epsilon_{\mu\rho} - \mathcal{P}_{\nu\rho} \epsilon_{\mu\sigma} \right)
\, . \\[1ex]
\end{split}
\end{equation}
Here, $\epsilon_{\mu\nu}$ is the polarization tensor of $2^{++}$ and $\mathcal{P}_{\mu\nu} \equiv g_{\mu\nu} - 2 p_\mu p_\nu /p^2$.
The results of the decay rates are given by
\begin{equation}
\begin{split}
\\[-2.5ex]
&\Gamma (2^{++} \rightarrow gg) = \frac{8 \alpha_s^2 \alpha_H^2}{160 \pi m_{X'}^8} \kappa_s^2 m_{2^{++}}^3
\left( \frac{1}{2} a_T^2 \left({\bf F^T_{2^{++}}} \right)^2 + \frac{4}{3} a_L^2 \left({\bf F^L_{2^{++}}} \right)^2 \right) \, , \\[2ex]
&\Gamma (2^{++} \rightarrow \gamma \gamma) = \frac{4 \alpha^2 \alpha_H^2}{160 \pi m_{X'}^8} \kappa_Y^2 m_{2^{++}}^3
\left( \frac{1}{2} a_T^2 \left({\bf F^T_{2^{++}}} \right)^2 + \frac{4}{3} a_L^2 \left({\bf F^L_{2^{++}}} \right)^2 \right)
\, , \\[2ex]
&\Gamma (2^{++} \rightarrow ZZ) = \frac{\alpha^2 \alpha_H^2\tan^4 \theta_W }{40 \pi m_{X'}^8} \kappa_Y^2 m_{2^{++}}^3
\left( 1 -4 x_2 \right)^{1/2} \biggl( \frac{1}{2} a_T^2 f_T (x_2)  \left({\bf F^T_{2^{++}}} \right)^2 \\[1ex]
&\qquad \qquad \qquad \qquad \qquad \qquad  + \frac{4}{3} a_L^2 f_L (x_2) \left({\bf F^L_{2^{++}}} \right)^2 
+ \frac{40}{3} a_T a_L f_{TL} (x_2) \, {\bf F^T_{2^{++}}} {\bf F^L_{2^{++}}}  \biggr)
\, , \nonumber \\[1ex]
\end{split}
\end{equation}
\begin{equation}
\begin{split}
\\[-2.5ex]
&\Gamma (2^{++} \rightarrow Z \gamma) = \frac{\alpha^2 \alpha_H^2\tan^2 \theta_W }{20 \pi m_{X'}^8} \kappa_Y^2m_{2^{++}}^3
\left( 1 -x_2 \right)^{3} \biggl( \frac{1}{2} a_T^2 g_T (x_2)  \left({\bf F^T_{2^{++}}} \right)^2 \\[1ex]
&\qquad \qquad \qquad \qquad \qquad \qquad  + \frac{4}{3} a_L^2 g_L (x_2) \left({\bf F^L_{2^{++}}} \right)^2 
+ \frac{20}{3} a_T a_L x_2 \, {\bf F^T_{2^{++}}} {\bf F^L_{2^{++}}}  \biggr)
\, , \\[1ex]
\end{split}
\end{equation}
where $x_2 = m_Z^2 / m_{2^{++}}^2$ and 
\begin{equation}
\begin{split}
\\[-2.5ex]
&f_T (x) = 1 - 3 x + 6 x^2 \, , \quad f_L (x) = 1 + 2 x + 36 x^2 \, , \quad f_{TL} (x) = x (1-x) \, , \\[1ex]
&g_T (x) = 1 + \frac{1}{2} x + \frac{1}{6} x^2 \, , \quad g_L (x) = 1 + 3 x + 6 x^2 
\, . \\[1ex]
\end{split}
\end{equation}
When we fix the scalar glueball mass at $m_0 = 750 \, \rm GeV$,
the mass of the spin 2 glueball $m_{2^{++}}$ is around $1 \, \rm TeV $.

\section{The $X'$ decay\label{sec:xdecay}}

\renewcommand{\arraystretch}{1.3}
\begin{table}[!t]
\begin{center}
\vspace{0.3cm}
\begin{tabular}{c|ccccc}
 & $SU(N+3)_{H}$ & $SU(3)_{C'}$ & $SU(2)_{L}$ & $U(1)_{Y'}$ 
 \\
 \hline
  $\psi$  &  $\mathbf{N+3}$ & $\mathbf {1}$ & $\mathbf  1$ & $\frac{3q}{N+3} $ \\
\end{tabular}
\end{center}
\caption{The charge assignment of the fermion relevant to the $X'$ decay.}
\label{tab:sunmodel2}
\end{table}
\renewcommand{\arraystretch}{1}

\renewcommand{\arraystretch}{1.3}
\begin{table}[!t]
\begin{center}
\vspace{0.3cm}
\begin{tabular}{c|ccccc}
 & $SU(N)_{H}$ & $SU(3)_{C}$ & $SU(2)_{L}$ & $U(1)_{Y}$ 
 \\
 \hline
  $\chi$  &  $\mathbf{{N}}$ & $\mathbf 1$ & $\mathbf  1$ & $0$ \\
  $\eta$ & $\mathbf{{1}}$ & $\mathbf{{3}}$ & $\mathbf  1$ & $q$  \\
\end{tabular}
\end{center}
\caption{The charge assignment of the fermion $\psi = (\chi, \eta)$ after the symmetry breaking.}
\label{tab:fermion}
\end{table}
\renewcommand{\arraystretch}{1}

In the simplest version of the model that we have discussed above, 
there is an unbroken $U(1)$ global symmetry under which 
the $X', \bar{X}'$ gauge bosons are charged 
and hence these massive gauge bosons are stable.
While this is not obviously ruled out experimentally and cosmologically
(at least if the reheating temperature after inflation is sufficiently low
and also non-thermal production of the $X', \bar{X}'$ gauge bosons is suppressed),
we here comment on a modest extension of the model
which allows the $X'$ boson to decay without breaking the $U(1)$ global symmetry.

Let us introduce a Dirac fermion $\psi$ charged under the $U(1)_{Y'}$ gauge group
and the $SU(N+3)_H$ gauge group with a Dirac
mass between $375$~GeV and $m_{Z'}/2$.
Table~\ref{tab:sunmodel2} and Table~\ref{tab:fermion} show the charge assignments of this fermion
$\psi = (\chi, \eta)$ before and after the symmetry breaking.  The $\chi$
and $\eta$ components are approximately degenerate because there are no
renormalizable couplings of $\psi$ to the $\xi$ field.
We assume that the new fermion interacts with the standard model matter fields.
The possible interaction depends on the charge $q$ of the $X'$ gauge boson.
For instance, when the charge is $q =5/3$, we can write down (for example)
the following invariant non-renormalizable
interaction term (in Majorana notation):
\begin{equation}
\frac{1}{M_{UV}^3}(u^{j_3}_R\beta u^{k_3}_R)
(d^c_{j_3L}\beta \psi^{c}_{j_NL})\xi^{j_N}_{k_3}
+\mbox{h.c.}
\label{Lpsi}
\end{equation}
where $u_R$ and $d_R$ are the ordinary right-handed up and down quarks, $j_3$
and $k_3$ are color indices and $j_N$ is an $SU(N)_H$ index, all summed over.
Then, if the $X'$ gauge boson is heavier than the $\chi$, it
can decay as follows:
\begin{equation}
X'\to \bar\chi uu\bar d \, .
\label{xdecay5over3}
\end{equation}
The electrically neutral fermion $\chi$ is stable and might be a candidate
of the dark matter.

The colored fermion $\eta$ can be produced at the LHC
but its collider phenomenology significantly depends on its charge and the
details of its decays. For example, for $q=5/3$, $\eta$ is a charge $5/3$
quark and the interaction term (\ref{Lpsi}) along with the $\xi$  VEV
produces the decay
\begin{equation}
\eta\to uu\bar d \, .
\label{etadecay5over3}
\end{equation}
If the lifetime of this fermion is long enough, the pair produced $\eta\bar\eta$
may form a bound state like charmonium.
If the decay (\ref{etadecay5over3}) is fast, 
we may see the $uu\bar d$ jets in the LHC detectors.
The detailed analysis is beyond the scope of this paper and will be
discussed elsewhere. 

If the scale $M_{UV}$ is very low, we may worry that the UV completion will
include flavor-changing netural-current effects.  It is interesting to note
that we can generalize (\ref{Lpsi}) to incorporate a $SU(3)_U\times
SU(3)_D$ symmetry acting on the right handed charge 2/3 and charge $-1/3$
quarks respectively.  The generalization, now including $SU(3)_U$ 
flavor indices, $j_U$, $k_U$ and $\ell_U$ and a $SU(3)_D$ flavor index
$j_D$ (again all summed over), looks like
\begin{equation}
\frac{1}{M_{UV}^3}(U^{j_3j_U}_R\beta U^{k_3k_U}_R)
\epsilon_{j_Uk_U\ell_U}
(D^c_{j_3j_DL}\beta \psi^{c\ell_Uj_D}_{j_NL})\xi^{j_N}_{k_3}
+\mbox{h.c.}
\label{Lpsiflavor}
\end{equation}
Now we have nine $\psi$ fields which carry the $U$ and $D$ flavor
symmetries (and it is amusing to note that this is getting close to the
number of $SU(N)$ fermions necessary slow the running of the $SU(N)$
coupling below the $m_{X'}$ scale).
Thus we can tune the coupling to preserve the flavor
symmetry. Likewise, we can adjust things so that the Dirac mass terms for
the $\psi$ fermions are equal, preserving the symmetry.  None of this is
natural but it suggests that the flavor changing neutral currents will not
be an insurmountable constraint, even if the coupling is fairly
strong.

\section{Conclusion\label{sec:conclusion}}

In this paper, we have described a partial unification model that
could explain the reported diphoton  
event excess.
A part of the color $SU(3)$ and the hypercharge $U(1)$ resides in an
extended gauge group that is broken by a VEV {\bf slightly smaller
  than the Higgs VEV!}   
We have discussed the experimental constraints on the new parameters.
Precise tests of the standard model at low energies
constrain the model parameters and require the coupling of the new gauge
group to be large.
Constraints from searches for
the $Z'$ and the coloron require that they decay dominantly into new particles.
The scalar glueball associated with the new confining gauge theory can have
a mass of around $750 \, \rm GeV$ and 
be produced by gluon fusion and decay into two photons through
loops of the new massive vector bosons $X, \bar{X}'$.
The production and decays are analyzed by the effective dimension eight
operator of the glueball 
and the mixing term with the singlet scalar.
We have found a parameter space where this glueball could explain the
diphoton excess. 
The decays of the pseudoscalar and spin $2$ glueballs have been also presented.

One of the important predictions in the present model is the existence of
the $X', \bar{X}'$ gauge bosons, 
which may be pair produced at colliders.
In the simplest version of the model, the $X', \bar{X}'$ gauge
bosons are stable.
We have discussed a modest extension of the
model which allows the $X'$ boson to decay into a colorless, neutral fermion.
The lifetime of the $X'$ gauge boson depends on the mass of the new fermion
and the size of the coupling of the interaction term with the standard model
field(s) like \eqref{Lpsi}. 
When the $X'$ gauge boson is stable at collider time scales,
the bound states of the $X', \bar{X'}$ gauge bosons, the vector bosoniums, are formed.
When the reported diphoton excess is explained in the present model,
the mass of the lightest vector bosonium is predicted at around 2 TeV.
Detailed phenomenology of the vector bosoniums is left for a future study.
It might be also interesting to clarify whether a stable baryonic bound
state of the neutral fermion could give 
the correct dark matter abundance.

Our model is neither natural nor beautiful, but we believe it is
instructive.  We close by reiterating a few of the things we have
noticed from the analysis that may be more generally useful.
\begin{itemize}
\item A partial unification not involving the electroweak $SU(2)$ can
  depend on an arbitrary charge, $q$ (see (\ref{qunify}) and the discussion following).
\item ``Flavor-diagonality'' is an inappropriate assumption for
  compositeness tests (see (\ref{coloroncontact}) and the discussion following).
\item Perhaps the most important and surprising message is that 
  a low partial unification scale with new particles that have
  large mass because their couplings to the symmetry breaking field are
  large may be only weakly constrained if the strong interactions do not
  directly involve the standard model fermions  (see (\ref{yparameter}) and
  the discussion following). 
\end{itemize}
Thus even if (as seems likely) the reported diphoton excess at $750$~GeV
is washed away in a flood of new data, we believe that we have learned
something. Our model is an explicit example of how new physics could
be hidden right in front of our noses at the $SU(2)\times U(1)$ breaking
scale and below. 

\section*{Acknowledgments\label{sec:ackno}}

We would like to thank Prateek Agrawal, Masaki Asano, Tatsuhiko Ikeda, Matthew Reece, Ryosuke Sato
and Matthew Strassler for fruitful discussions.
We are particularly grateful to M. Reece for many important comments from
the very beginning of this work.  HG is supported in part by the National
Science Foundation under grant PHY-1418114. 
YN is supported by a JSPS Fellowship for Research Abroad.

\appendix

\section{Normalization and identities of group theory\label{sec:identities}}

We here summarize normalization and identities of $SU(N+3)$ and its subgroups.
The commutation relations are
\begin{equation}
\begin{split}
\\[-2.5ex]
[T^A, T^B] = i f^{ABC} T^C \, , \\[1ex]
\end{split}
\end{equation}
where $T^A$ ($A = 1, \cdots , (N+3)^2 -1$) are generators and $f^{ABC}$ are totally antisymmetric structure constants.
The anti-commutation relations are
\begin{equation}
\begin{split}
\\[-2.5ex]
\{ T^A, T^B \} = \delta^{AB} \frac{\bf 1}{N+3} + d^{ABC} T^C \, , \\[1ex]
\end{split}
\end{equation}
where $d^{ABC}$ are totally symmetric.
There are relations,
\begin{equation}
\begin{split}
\\[-2.5ex]
\tr (T^A_R T^B_R) = C(R) \delta^{AB} \, , \qquad f^{ABC} = - \frac{i}{C(R)} \tr \left ([T^A_R , T^B_R] T^C \right) \, , \\[1ex]
\end{split}
\end{equation}
where $R$ denotes a representation.
For the (anti-)fundamental representations, $\bf N+3$ and $\bf \overline{N+3}$, $C(R) = 1/2$.
We also have
\begin{equation}
\begin{split}
\\[-2.5ex]
T^A_R T^A_R = C_2 (R) {\bf 1 } \, , \qquad f^{ABC} f^{ABD} = (N+3) \delta^{CD} \, , \\[1ex]
\end{split}
\end{equation}
where $C_2 (R)$ is the quadratic Casimir and $C_2(G) = N+3$ for the adjoint representation.

Next, let us divide the  $SU(N+3)$ generators into the generators of the subgroups $U(1) \times SU(3) \times SU(N)$ and
the other non-hermitian generators.
We denote the $SU(3)$ generators as
\begin{equation}
\begin{split}
\\[-2.5ex]
T^a \, , \, T^b \, , \qquad a, b = 1, \cdots , 8 \, , \\[1ex]
\end{split}
\end{equation}
which satisfy
\begin{equation}
\begin{split}
\\[-2.5ex]
[T^a, T^b] = i f^{abc} T^c \, , \qquad f^{abc} f^{abd} = 3 \delta^{cd}  \, , \\[1ex]
\end{split}
\end{equation}
and the $U(1)$ generator as $T^9 \equiv \tilde{S}$.
The $SU(N)$ generators are
\begin{equation}
\begin{split}
\\[-2.5ex]
T^m \, , \, T^n \, , \qquad m, n = 10 + 6N , \cdots , (N+3)^2 - 1 \, , \\[1ex]
\end{split}
\end{equation}
which satisfy
\begin{equation}
\begin{split}
\\[-2.5ex]
[T^m, T^n] = i f^{mnl} T^l \, , \qquad f^{mnl} f^{mnk} = N \delta^{lk}  \, , \\[1ex]
\end{split}
\end{equation}
and the $6N$ non-hermitian generators are
\begin{equation}
\begin{split}
\\[-2.5ex]
T^p \, , \, T^{\bar{q}} \, , \qquad p, \bar{q} = 10 , \cdots , 9+ 3N \, , \\[1ex]
\end{split}
\end{equation}
which satisfy
\begin{equation}
\begin{split}
\\[-2.5ex]
\tr \left( T^p T^{\bar{q}} \right) = \frac{1}{2} \delta^{pq} \, , \\[1ex]
\end{split}
\end{equation}
for the fundamental representations.
The commutation relations of the generators among the $SU(3)$, $U(1)$ and $SU(N)$ subgroups are zero,
\begin{equation}
\begin{split}
\\[-2.5ex]
[T^a, T^m] = [T^a, T^9] = [T^m, T^9] = 0 \, . \\[1ex]
\end{split}
\end{equation}
We also have
\begin{equation}
\begin{split}
\\[-2.5ex]
&\tr \left( T^a T^m \right) = \tr \left( T^a T^9 \right) = \tr \left( T^a T^p \right) = \tr \left( T^a T^{\bar{p}} \right)  \\[1ex]
&=\tr \left( T^m T^9 \right) = \tr \left( T^m T^p \right) = \tr \left( T^m T^{\bar{p}} \right)
= \tr \left( T^9 T^p \right) = \tr \left( T^9 T^{\bar{p}} \right) = 0 \, . \\[1ex]
\end{split}
\end{equation}
and
\begin{equation}
\begin{split}
\\[-2.5ex]
f^{pqa} = f^{\bar{p} \bar{q} a} = 0 \, . \\[1ex]
\end{split}
\end{equation}
Then, we can derive the following useful formula,
\begin{equation}
\begin{split}
\\[-2.5ex]
f^{p\bar q a} (f^{p\bar q b})^\ast
&=f^{\bar q p a} (f^{\bar q p b})^\ast \\[1ex]
&=\frac{1}{2}\left(f^{AB a}f^{AB b}
-f^{cd a}f^{cd b}\right) \\[1ex]
&= \frac{1}{2}\left((N+3)-3\right)\delta^{ab}=\frac{N}{2}\delta^{ab} \, . \\[1ex]
\end{split}
\end{equation}
Note that $(f^{p\bar q a})^\ast = f^{\bar p q a}$.
In the same way, we have
\begin{equation}
\begin{split}
\\[-2.5ex]
f^{p\bar q m} (f^{p\bar q n})^\ast
&=f^{\bar q p m} (f^{\bar q p n})^\ast \\[1ex]
&=\frac{1}{2}\left(f^{AB m}f^{AB n}
-f^{kl m}f^{kl n}\right) \\[1ex]
&= \frac{1}{2}\left((N+3)-N\right)\delta^{mn}=\frac{3}{2}\delta^{mn} \, . \\[1ex]
\end{split}
\end{equation}
Note that $(f^{p\bar q m})^\ast = f^{\bar p q m}$.

\section{Summary of interactions\label{sec:interactions}}

In this appendix, we summarize some of the interactions in the model.

\subsection{Coloron interactions}\label{coloronintsummary}

We concentrate on the coloron interactions relevant to the coloron decays.
The interaction which leads to $G' \rightarrow G_O G_O$ is given by
\begin{equation}
\begin{split}
\\[-2.5ex]
\mathcal{L} &\supset  \frac{g_H^2 -  (g'_3)^2}{\sqrt{g_H^2 +  (g'_3)^2}} f^{abc} (\partial^\mu G_O^a) G{^\prime}^b_\mu G_O^c  \, . \\[1ex]
\end{split}
\end{equation}
The relevant interaction of $G' \rightarrow G_O Z'$ is
\begin{equation}
\begin{split}
\\[-2.5ex]
\mathcal{L} \supset - \sqrt{\frac{2}{3}} \, m_{Z'} \sqrt{g_H^2 + (g'_3)^2} \, G_O^a Z'_\mu G{^\prime}^{\mu a } \, .  \\[1ex]
\end{split}
\end{equation}
The interactions to give $G' \rightarrow X'\bar{X}'$ are
\begin{equation}
\begin{split}
\\[-2.5ex]
\mathcal{L} \supset \frac{g_H^2}{\sqrt{g_H^2 + (g'_3)^2}} \, f^{ap\bar{q}} \,
\Bigl\{ &- (\partial_\kappa G{^\prime}^a_\lambda) X{^\prime}^{\kappa p} \bar{X}{^\prime}^{\lambda \bar{q}}
+ (\partial_\kappa {X}{^\prime}^{p}_\lambda ) G{^\prime}^{\kappa a} \bar{X}{^\prime}^{\lambda \bar{q}}
- (\partial_\kappa {X}{^\prime}^{p}_\lambda ) \bar{X}{^\prime}^{\kappa \bar{q}} G{^\prime}^{\lambda a}   \\[1ex]
&+ (\partial_\kappa G{^\prime}^a_\lambda) \bar{X}{^\prime}^{\kappa \bar{q}} {X}{^\prime}^{\lambda p} 
- (\partial_\kappa \bar{X}{^\prime}^{\bar{q}}_\lambda ) G{^\prime}^{\kappa a} {X}{^\prime}^{\lambda p}
+ (\partial_\kappa \bar{X}{^\prime}^{\bar{q}}_\lambda ) {X}{^\prime}^{\kappa p} G{^\prime}^{\lambda a}  \Bigr\} \, . \\[1ex]
\end{split}
\end{equation}

\subsection{New massive gauge boson interactions}\label{XXScalar}

We here present the $X', \bar{X}'$ interactions which lead to the mixing between the scalar $N$-glueball and the singlet scalar $G_S$.
The $X', \bar{X}'$ gauge boson interaction with the scalar $G_S$ is given by
\begin{equation}
\begin{split}
\\[-2.5ex]
\mathcal{L} &\supset
\frac{1}{\sqrt{6}} \, g_H^2 a \, G_S X'_\mu \bar{X}{^\prime}^\mu  \, . \\[1ex]
\end{split}
\end{equation}
The cubic interactions of the $X', \bar{X}'$ gauge bosons with the $SU(N)_H$ gauge field $H_\mu^m$ are
\begin{equation}
\begin{split}
\\[-2.5ex]
\mathcal{L} \supset &- g_H f^{mp\bar{q}} (\partial_\kappa {H}_\lambda^m) {X}{^\prime}^{\kappa p} \bar{X}{^\prime}^{\lambda \bar{q}}
+ g_H f^{m p \bar{q}} (\partial_\kappa {X}{^\prime}^{p}_\lambda ) {H}^{\kappa m} \bar{X}{^\prime}^{\lambda \bar{q}}
- g_H f^{m p \bar{q}} (\partial_\kappa {X}{^\prime}^{p}_\lambda ) \bar{X}{^\prime}^{\kappa \bar{q}} {H}^{\lambda m}   \\[1ex]
&+ g_H f^{mp\bar{q}} (\partial_\kappa {H}_\lambda^m) \bar{X}{^\prime}^{\kappa \bar{q}} {X}{^\prime}^{\lambda p} 
- g_H f^{m p \bar{q}} (\partial_\kappa \bar{X}{^\prime}^{\bar{q}}_\lambda ) {H}^{\kappa m} {X}{^\prime}^{\lambda p}
+ g_H f^{m p \bar{q}} (\partial_\kappa \bar{X}{^\prime}^{\bar{q}}_\lambda ) {X}{^\prime}^{\kappa p} {H}^{\lambda m}  \, . \\[1ex]
\end{split}
\end{equation}
The quartic interactions are given by
\begin{equation}
\begin{split}
\\[-2.5ex]
\mathcal{L} \supset
&- g_H^2 ( f^{mnl} {H}_\kappa^m {H}_\lambda^n) ( f^{p \bar{q}l} X{^\prime}^{\kappa p} \bar{X}{^\prime}^{\lambda \bar{q}})  \\[1ex]
&- g_H^2 ( f^{rm\bar q} {H}_\kappa^m \bar{X}{^\prime}^{\bar{q}}_\lambda) ( f^{\bar{r} n p} H^{\kappa n} X{^\prime}^{\lambda p} ) \\[1ex]
&+ g_H^2 ( f^{rm\bar q} {H}_\kappa^m \bar{X}{^\prime}^{\bar{q}}_\lambda) ( f^{\bar{r} n p}  X{^\prime}^{\kappa p} H^{\lambda n} ) \, . \\[1ex]
\end{split}
\end{equation}

\section{The effective operator coefficients\label{sec:dimeight}}

We here identify the coefficients of the effective dimension eight operators presented in the main text,
$a_S$, $a_P$, $a_T$, $a_L$ and $b_1$, $b_2$.
First, we have the relation,
\begin{equation}
\begin{split}
\\[-2.5ex]
\epsilon_{\mu\nu\rho\sigma} \epsilon_{\alpha\beta\gamma\delta} 
= - g_{\alpha \zeta} g_{\beta \eta} g_{\gamma \theta} g_{\delta \xi} 
\det \begin{pmatrix}
\delta^{\zeta}_{\rho} & \delta^{\zeta}_{\sigma} & \delta^{\zeta}_{\mu} & \delta^{\zeta}_{\nu} \\
\delta^{\eta}_{\rho} & \delta^{\eta}_{\sigma} & \delta^{\eta}_{\mu} & \delta^{\eta}_{\nu} \\
\delta^{\theta}_{\rho} & \delta^{\theta}_{\sigma} & \delta^{\theta}_{\mu} & \delta^{\theta}_{\nu} \\
\delta^{\xi}_{\rho} & \delta^{\xi}_{\sigma} & \delta^{\xi}_{\mu} & \delta^{\xi}_{\nu}
\end{pmatrix}
\, . \\[1ex]
\end{split}
\end{equation}
Using this relation, the effective Lagrangian \eqref{dimeight} can be rewritten as
\begin{equation}
\begin{split}
\\[-2.5ex]
\mathcal{L}_{\rm eff} = \, &\frac{g_H^2}{(4\pi)^2 m_{X'}^4} \biggl\{ \\[1ex]
&\left( g_Y^2 \chi_Y B_{\rho \sigma} B^{\rho\sigma}
+ g_s^2 \chi_s \tr \,  G_{\rho \sigma} G^{\rho\sigma}  \right) \left( a_S - \frac{1}{4} a_T + \frac{1}{3} a_L \right)
\tr \, F_{\alpha \beta} F^{\alpha \beta}  \\[1ex]
&+\left( g_Y^2 \chi_Y B^{\mu\nu} B^{\rho\sigma}
+ g_s^2 \chi_s \tr \,  G^{\mu\nu} G^{\rho\sigma}  \right) \left( 8 a_P + \frac{2}{3} a_L  \right)
\tr \, F_{\rho \mu} F_{\sigma \nu}  \\[1ex]
&-\left( g_Y^2 \chi_Y B^{\mu\nu} B^{\rho\sigma}
+ g_s^2 \chi_s \tr \,  G^{\mu\nu} G^{\rho\sigma}  \right) \left( 4 a_P - \frac{2}{3} a_L  \right)
\tr \, F_{\rho \sigma} F_{\mu \nu}  \\[1ex]
&+\left( g_Y^2 \chi_Y B^{\mu}_{\,\, \sigma} B^{\rho\sigma}
+ g_s^2 \chi_s \tr \,  G^{\mu}_{\,\, \sigma} G^{\rho\sigma}  \right) \left( a_T - 2 a_L  \right)
\tr \, F_{\mu \lambda} F_{\rho}^{\,\, \lambda} \biggr\} \\[1ex]
&+ \frac{g_H^3 g_Y}{(4\pi)^2 m_{X'}^4} \kappa_\Omega \left( b_1 B^{\mu\nu} \Omega^{(1)}_{\mu\nu}
+ b_2 B^{\mu\nu} \Omega^{(2)}_{\mu\nu}  \right) \, . \\[1ex]
\end{split}
\end{equation}
The general expression of the effective Lagrangian has been calculated in \cite{Metsaev:1987ju,Fichet:2013ola}.
Using this result, we obtain
\begin{equation}
\begin{split}
\\[-2.5ex]
&a_S = \frac{7}{288} (\gamma_1 + \gamma_2)  + \frac{1}{18} (\gamma_3 + \gamma_4)  \, ,
\qquad a_P = \frac{1}{288} (\gamma_1 + \gamma_2)  - \frac{1}{144} (\gamma_3 + \gamma_4)  \, , \\[1ex]
&a_T = \frac{1}{8} (\gamma_1 + \gamma_2)  + \frac{1}{6} (\gamma_3 + \gamma_4)  \, ,
\qquad a_L = \frac{1}{48} (\gamma_1 + \gamma_2)  + \frac{1}{12} (\gamma_3 + \gamma_4)  \, , \\[1ex]
&b_1 = \frac{1}{12} (\gamma_3 + \gamma_4)  \, , \qquad b_2 = \frac{1}{12} (\gamma_1 + \gamma_2) \, , \\[1ex]
\end{split}
\end{equation}
where $(\gamma_1, \gamma_2, \gamma_3, \gamma_4) = \left( \frac{342}{35} , \frac{675}{105} , -\frac{621}{210}, - \frac{333}{420}\right)$
for a spin one particle integrated out.
Then, we have the coefficients presented in \eqref{coefficients}.

\bibliography{sunmodel}
\bibliographystyle{utphys}

\end{document}